\documentclass[12pt]{iopart}

\usepackage{graphicx}
\usepackage{subfigure}
\usepackage{lineno}
\usepackage{hyperref}
\usepackage{color}

\usepackage{caption}    
\usepackage{subcaption} 
\begin{document}
\title[A kinetic CMA diagram]{A kinetic CMA diagram}

\author{Zilong Li$^{1,2}$,Haotian Chen$^{2\dagger}$, Zhe Gao$^{1\dagger}$ and Wei Chen$^2$}

\address{$^1$ Department of Engineering Physics, Tsinghua University, Beijing 100084, People's Republic of China}
\address{$^2$ Southwestern Institute of Physics, P.O. Box 432, Chengdu 610041, People's Republic of China}

\ead{chenhaotian@swip.ac.cn, gaozhe@tsinghua.edu.cn}

\begin{abstract}
 
We present a kinetic Clemmow-Mullaly-Allis (CMA) diagram by systematically analysing the kinetic effects on the wave propagation in a homogeneous thermal plasma.
The differences between the cold and kinetic CMA diagrams are outlined.
It is found that new boundaries for weakly damped left- and right-handed circularly polarized waves are located above the ion and electron cyclotron frequency lines in the kinetic CMA diagram.
Additionally, Langmuir waves in the kinetic CMA diagram occupy a specific region between the new Langmuir wave boundary and the plasma frequency line, while in the cold CMA diagram, they exist on the plasma frequency line.
The extraordinary-Bernstein mode transformation frequency lines in the kinetic CMA diagram replace the hybrid resonant frequency lines of the cold CMA diagram, with discontinuities between different cyclotron harmonics.
These new boundaries partition the parameter space in the kinetic CMA diagram differently, leading to new inverse wave normal surfaces in the regions bounded by new boundaries.
The kinetic CMA diagram not only contributes to a basic understanding of wave properties in thermal plasmas, but also can provide a powerful tool to explore new possible propagation paths.

\end{abstract}

\vspace{2pc}

\section{Introduction}
Even cold plasma waves, the most basic plasma wave problem can become complicated when considering different parameter regimes.
In order to elucidate the characteristics of cold plasma waves, the Clemmow-Mullaly-Allis (CMA) diagram was proposed in 1950-60s \cite{CM1955,Allis1961}.
The fundamental theorems on the wave normal surface (WNS) established by Stix \cite{stix1992} demonstrate that the topological features of WNSs remain invariant within specified regimes and change only when crossing each boundary.
The CMA diagram is thus a standard tool for investigating the cut-off, resonance, and other characteristics of cold plasma waves in local plasma limit \cite{F.F.Chen, Bindslev2004}.
However, generalizing the CMA diagram to thermal plasmas faces essential challenges arising from kinetic effects \cite{swanson, krall1973, chenliu1987}. 

In this study, we present a kinetic CMA diagram for uniform thermal plasmas by self-consistently taking into account kinetic effects. 
The kinetic normal mode spectra are systematically calculated by using a novel algorithm based on the generalized argument principle \cite{chenhaotian2022}.
In particular, we will investigate (i) the differences between the cold and kinetic CMA diagrams, and (ii) the applications of the kinetic CMA diagram.
New boundaries are proposed for collisionless damped left- (L) and right-handed (R) circularly polarized waves, Langmuir and Bernstein waves. 
The kinetic CMA diagram offers a comprehensive framework for elucidating wave propagation paths across the entire frequency spectrum.
As an application, it is shown that the ion Bernstein wave in tokamak plasmas can propagate to the plasma edge ($\omega_{pe}^{2}/\omega^{2}=1$) if the parallel index of refraction ($n_{\|}$) is larger than the slow-fast wave mode conversion threshold.
In contrast, the ion Bernstein waves with $n_{\|}$ smaller than the threshold will convert to fast waves and are difficult to be measured outside the plasma. 
Meanwhile, it is found that the electron Bernstein waves are more easily detected in the high field side of tokamaks compared to low field side. 

This article is structured as follows.
In Section II, we revisit the CMA diagram of cold plasmas with the inverse wave normal surface (IWNS). 
Section III presents the kinetic CMA diagram.
Finally, Sec.IV states concluding remarks and discussions.

\section{The CMA Diagram with Inverse Wave Normal Surfaces}
We first revisit the CMA diagram in cold plasmas with IWNSs.
Considering a plasma composed of electrons and hydrogen ions (${M}/{m}=1836$) as an illustrative example, the cold plasma dispersion relation is given by \cite{stix1992}
\begin{equation}
  tan^{2}\theta= \frac{P(n^{2}-R)(n^{2}-L)}{(RL-Sn^{2})(n^{2}-P)},
  \label{eq:cold_dispersion_relation}
\end{equation}
where $\theta$ is the angle between the wave vector $\mathbf{k}$ and magnetic field $\mathbf{B}$, and $n={kc}/{\omega},R=S+D,L=S-D$,
$S=1+\sum_{j=i,e}{\omega_{pj}^{2}}/{(\Omega_{cj}^{2}-\omega^{2})},D=-\sum_{j=i,e}{\Omega_{cj}\omega_{pj}^{2}}/{\omega(\Omega_{cj}^{2}-\omega^{2})},
P=1-\sum_{j=i,e}{\omega_{pj}^{2}}/{\omega^{2}},$
with the cyclotron frequencies $\Omega_{ci}={eB}/{M},\Omega_{ce}=-{eB}/{m}$.
By taking the limits $n^{2}\rightarrow 0$ and $n^{2}\rightarrow +\infty$ in Eq.(\ref{eq:cold_dispersion_relation}) respectively, we can derive the cutoff and resonant lines in the CMA diagram: (1) the L cutoff line, $L=0$; (2) the R cutoff line, $R=0$; (3) the P cutoff line, $P=0$; (4) the lower and upper hybrid resonant lines, $P+Stan^{2}\theta=0$, where $S=0$ for the case when $\theta=\pi/2$. 

\begin{figure}
  \centering
  \includegraphics[width=0.6\textwidth]{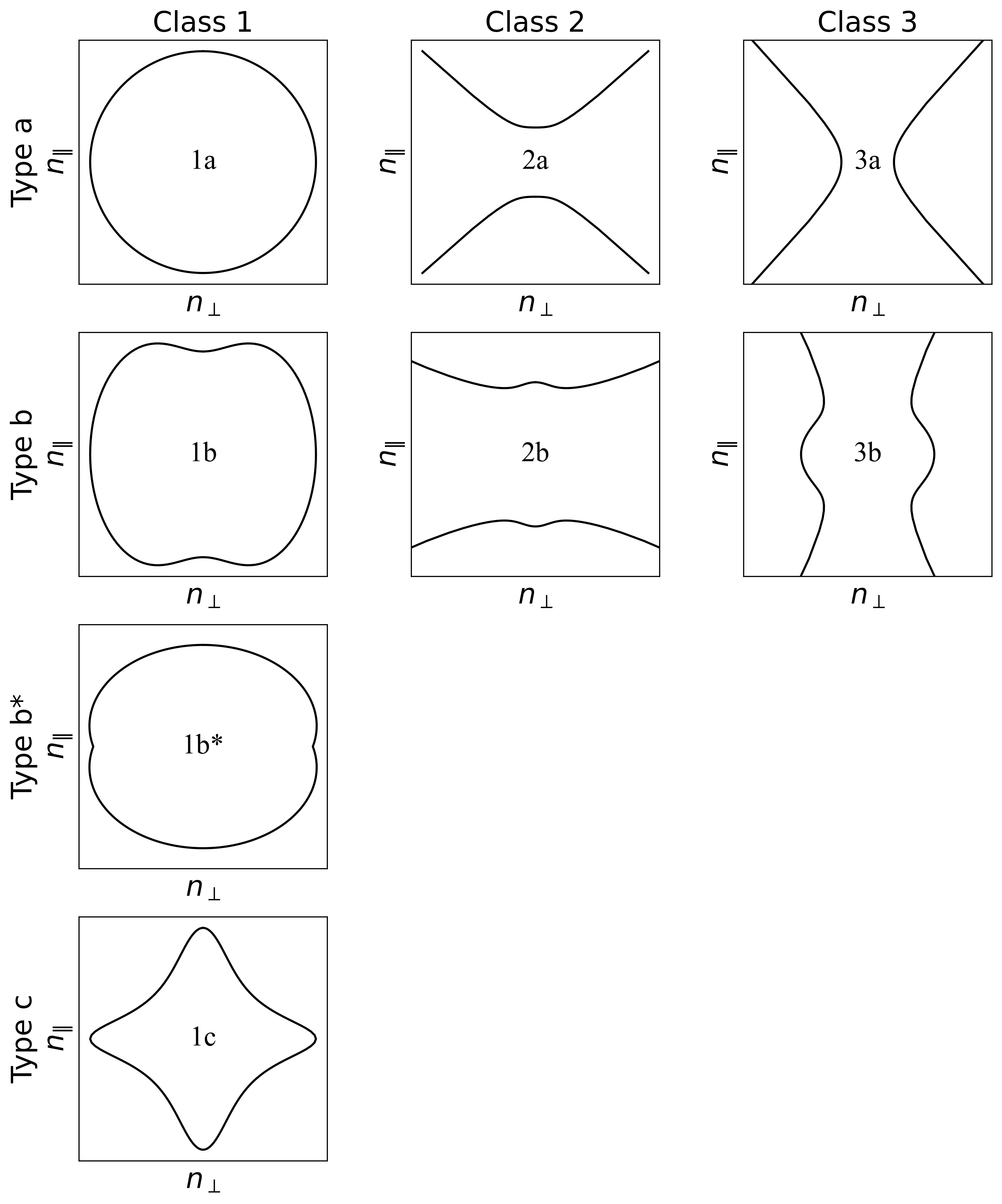}
  \caption{The eight types of IWNSs in the cold plasma CMA diagram.\label{fig:iwns class}}
\end{figure}

In this paper, we have adopted the IWNSs rather than the WNSs in the CMA diagram \cite{Takahashi1994,Pinsker2020}.
This choice stems from a crucial characteristic of IWNSs that the group velocity at any given point of the IWNS is perpendicular to the local tangent \cite{stix1992}.
Given that the parallel wavenumber $k_{\|}$ is typically regarded as constant during wave propagation through the axisymmetric tokamak plasma, the use of IWNSs becomes particularly well-suited for delineating wave propagation paths characterized by a fixed, finite $k_{\|}$.
As illustrated in Fig.\ref{fig:iwns class}, there are eight types of IWNSs in the CMA diagram \cite{stott, walker1, walker2, walker3} while the IWNS boundaries are summarized in table 1. The Appendix A provides a brief derivation of the IWNS boundaries.
\begin{table}
  \caption{\label{table1}Summary of IWNS boundaries (see Appendix A for more details). }
  \begin{indented}
  \item[]\begin{tabular}{cccc}
  \br
  Wave mode & Boundary equation & Constraint & IWNS transition\\ 
  \mr
  L wave & $P+L=0$ & $P<0$ & Type 1a and 1b; Type 2a and 2b \\ 
  R wave & $P+R=0$ & $P<0$ & Type 1a and 1b; Type 2a and 2b \\ 
  X wave & $(P+S)RL-2PS^{2}=0$ & $RL/S>0$ & Type 1a and 1b*; Type 3a and 3b \\ 
  O wave & $P-S=0$ & $P>0$ & Type 1a and 1b* \\ 
  \br
  \end{tabular}
  \end{indented}
\end{table}

\begin{figure}
  \centering
  \subfigure[Overall figure]
  {
    \begin{minipage}[b]{0.64\textwidth}
      \centering
      \includegraphics[width=\textwidth]{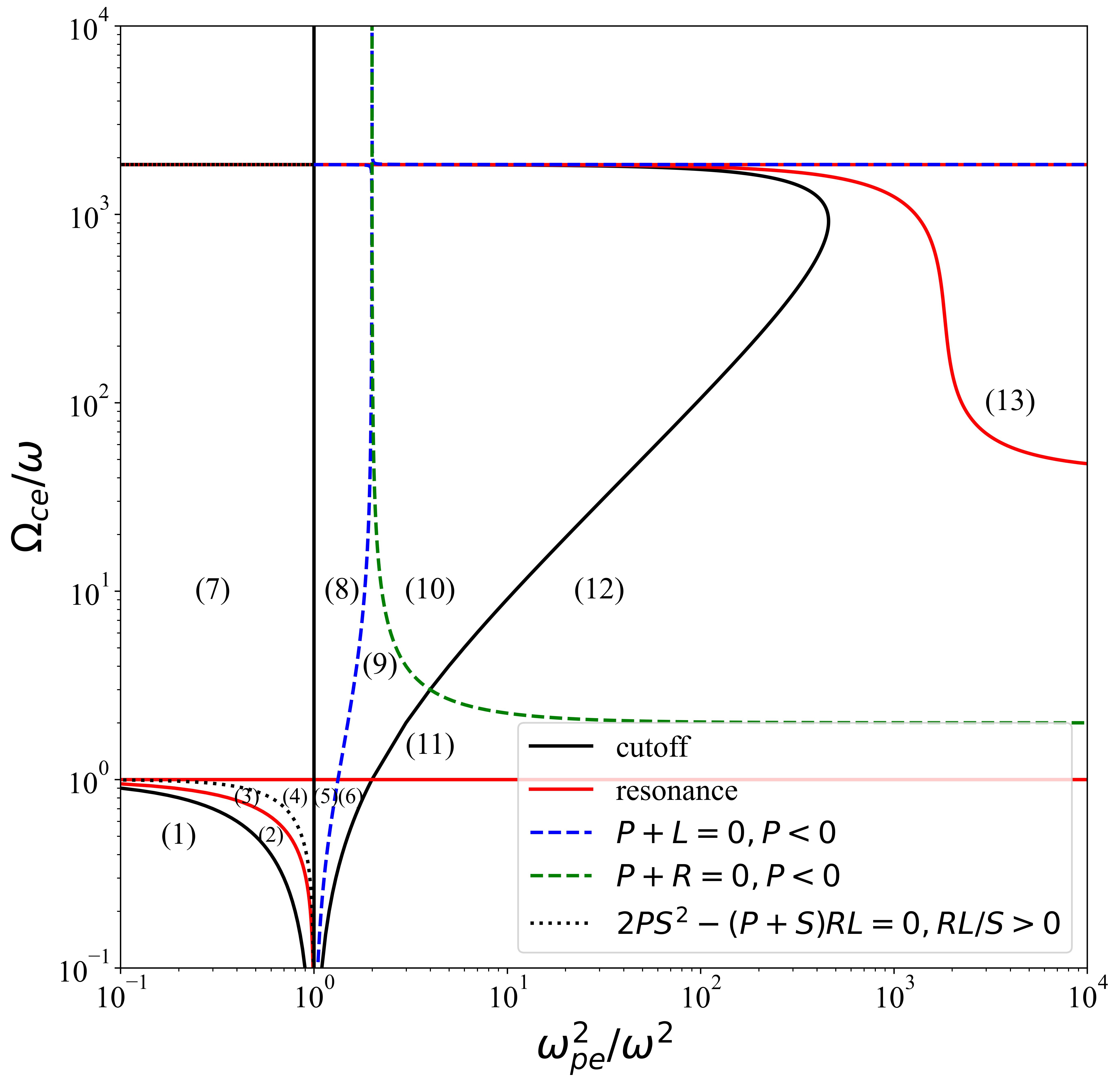}
    \end{minipage}
  }
  \subfigure[Enlarged view of the low-frequency regime]
  {
    \begin{minipage}[b]{0.64\textwidth}
      \centering
      \includegraphics[width=\textwidth]{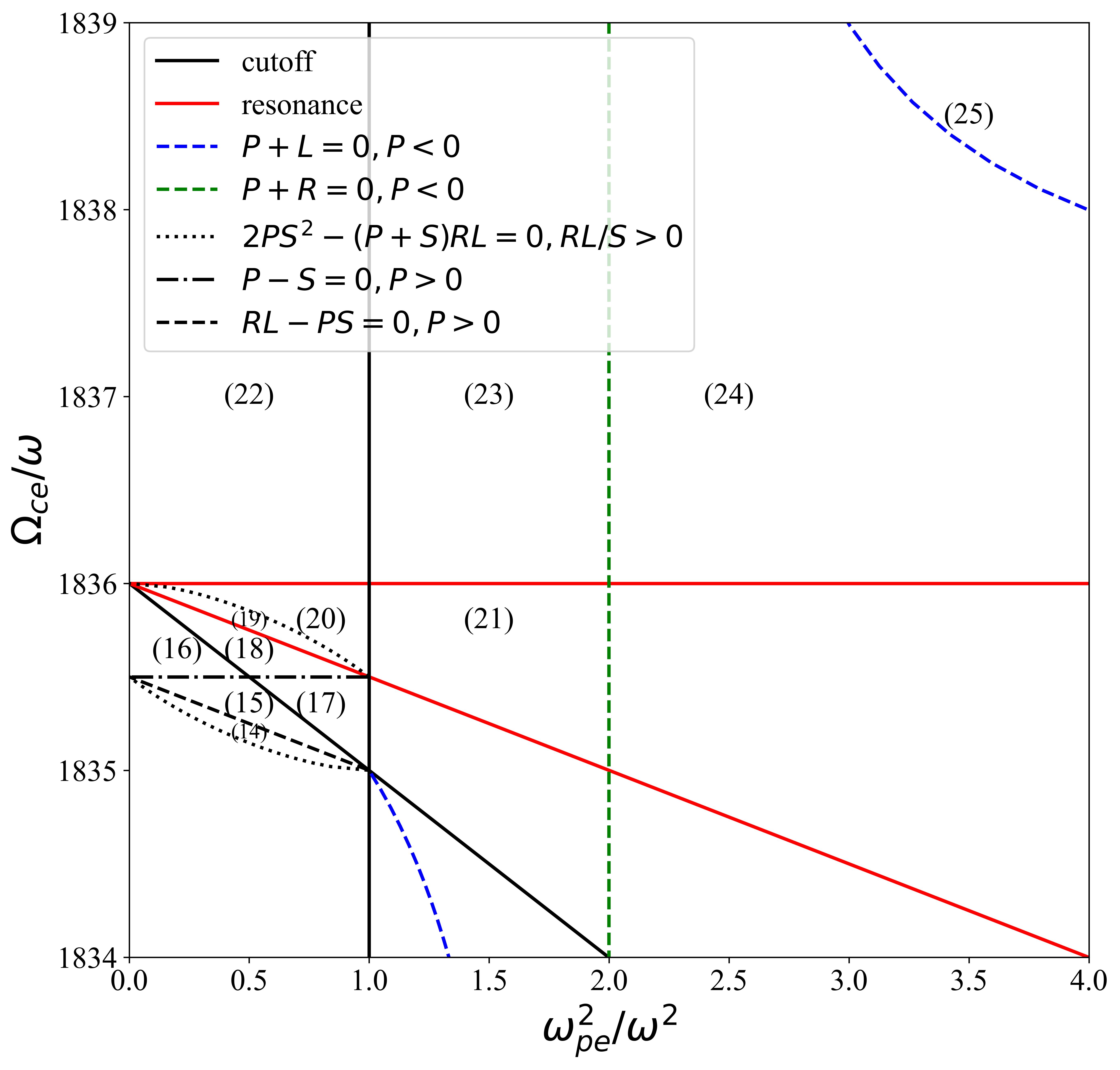}
    \end{minipage}
  }
  \caption{\label{fig:cold cma diagram} The CMA diagram with the IWNS boundaries for both high frequency and low frequency regimes. The black and red solid lines represent cutoff and resonant lines, respectively, while the IWNS boundaries are displayed by the dashed lines.}
\end{figure}

The CMA diagram including the boundaries of IWNSs is shown in Fig.\ref{fig:cold cma diagram} and the associated IWNSs for each region are illustrated in Fig.\ref{fig:cold iwns surface}.
In this context, the ordinary mode refers to the wave propagating perpendicular to $\mathbf{B}_{0}$ with $\mathbf{E}_{1}$ parallel to $\mathbf{B}_{0}$, whereas the extraordinary mode describes the wave that propagates perpendicular to $\mathbf{B}_{0}$ with $\mathbf{E}_{1}$ perpendicular to $\mathbf{B}_{0}$.
It is important to note that the cutoff lines in the CMA diagram are independent of the propagation angle, whereas the resonant lines in Fig.\ref{fig:cold cma diagram} are merely applicable to parallel and perpendicular waves.  
Furthermore, as the kinetic dispersion relation converges to the cold plasma limit when the wavenumber approaches toward zero \cite{stix1992,swanson},  the kinetic and cold CMA diagrams share the same cutoff lines.
However, the resonant lines depend on kinetic effects. 
These points will be discussed in detail in Section 3.

\begin{figure}
  \centering
  \includegraphics[width=0.81\textwidth]{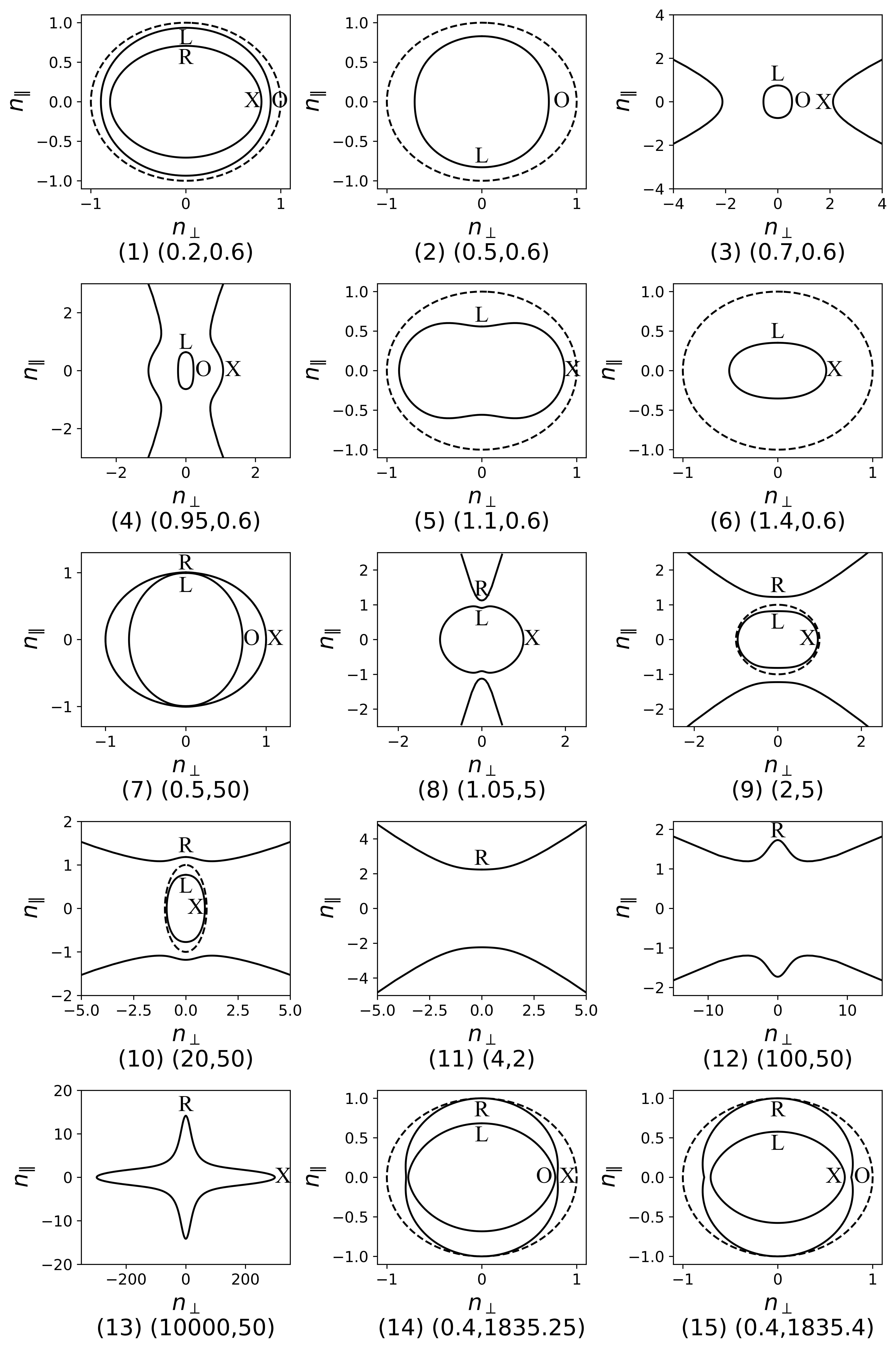} 
  \caption*{(1)-(15)} 
  \label{fig:subfig1}
\end{figure}

\begin{figure}
  \centering
  \includegraphics[width=0.81\textwidth]{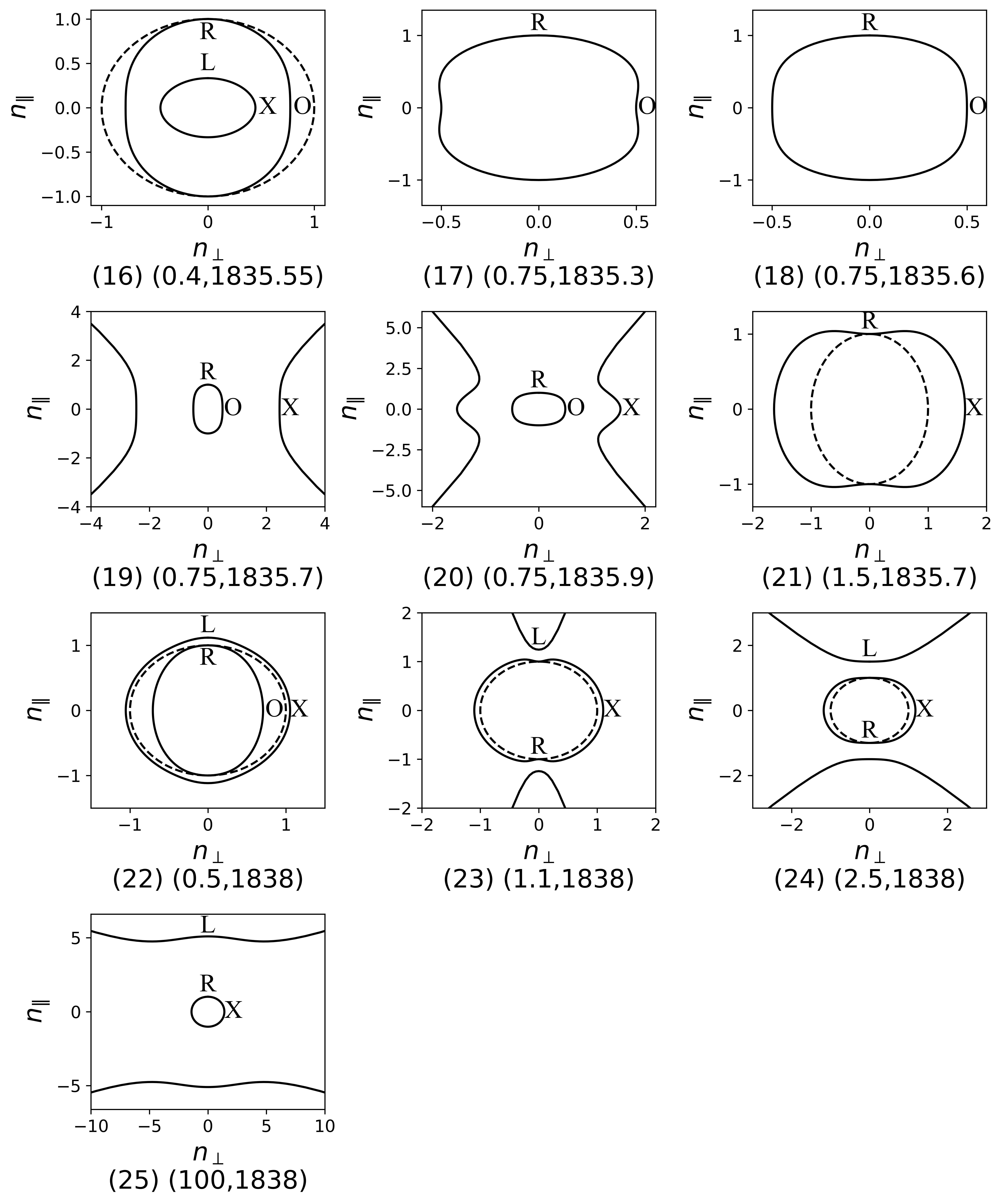} 
  \caption*{(16)-(25)} 
  \label{fig:subfig2}
\end{figure}

\begin{figure}
  \centering
  \caption{The representative IWNSs in each region of Fig.\ref{fig:cold cma diagram}. The dashed line represents $n^{2}=1$ surface. The coordinate below each subfigure indicates the position in the CMA diagram.}
  \label{fig:cold iwns surface}
\end{figure}

The CMA diagram serves as a fundamental framework for analysing wave propagation across the full frequency spectrum in uniform plasma. 
Although the concept is similar to the present homogeneous plasma analyses \cite{swanson}, discussions about wave propagation paths that take into account variations in the parallel wavenumber are intrinsically more complex.
As the current research is aimed at proposing a kinetic CMA diagram for homogeneous plasmas, we adopt the approximation that the parallel wavenumber $k_{\|}$ remains constant along the propagation path in axisymmetric tokamak plasmas. 
This simplification allows us to focus on the development of the kinetic CMA diagram without the added complexity of variable $k_{\|}$ values, which would otherwise complicate the analysis significantly.
Below, we outline an illustrative case of mode transformation under this approximation.

\begin{figure*}
  \centering
  \subfigure[$n_{\|}=2.0$]
  {
    \begin{minipage}[b]{1.0\textwidth}
      \centering
      \includegraphics[width=\textwidth]{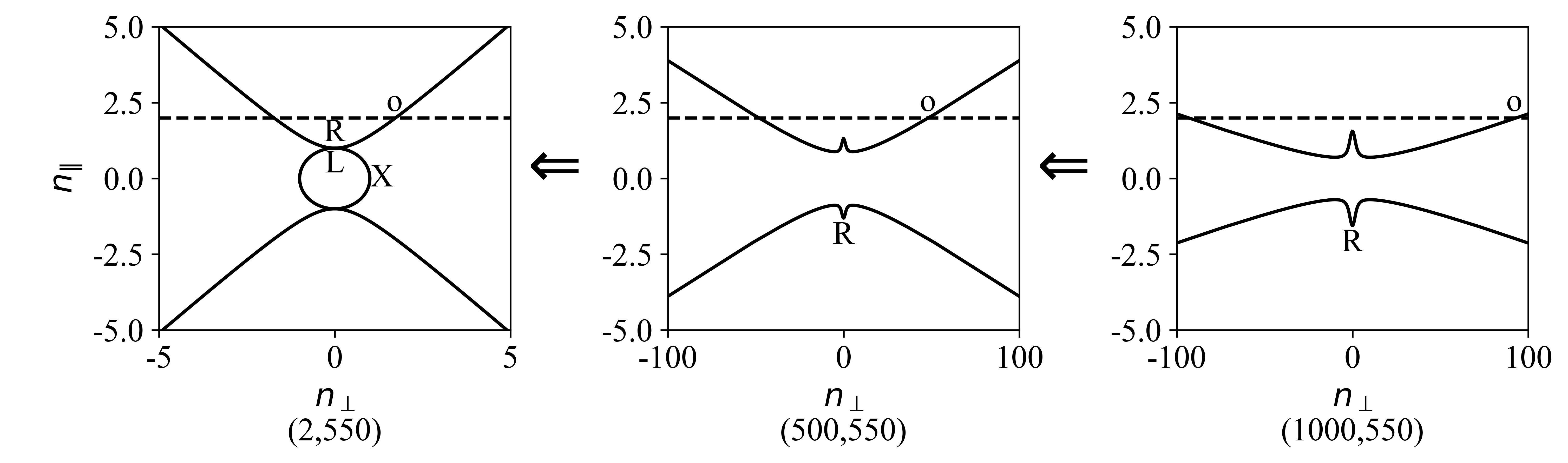}
    \end{minipage}
  }
  \subfigure[$n_{\|}=0.64$]
  {
    \begin{minipage}[b]{1.0\textwidth}
      \centering
      \includegraphics[width=\textwidth]{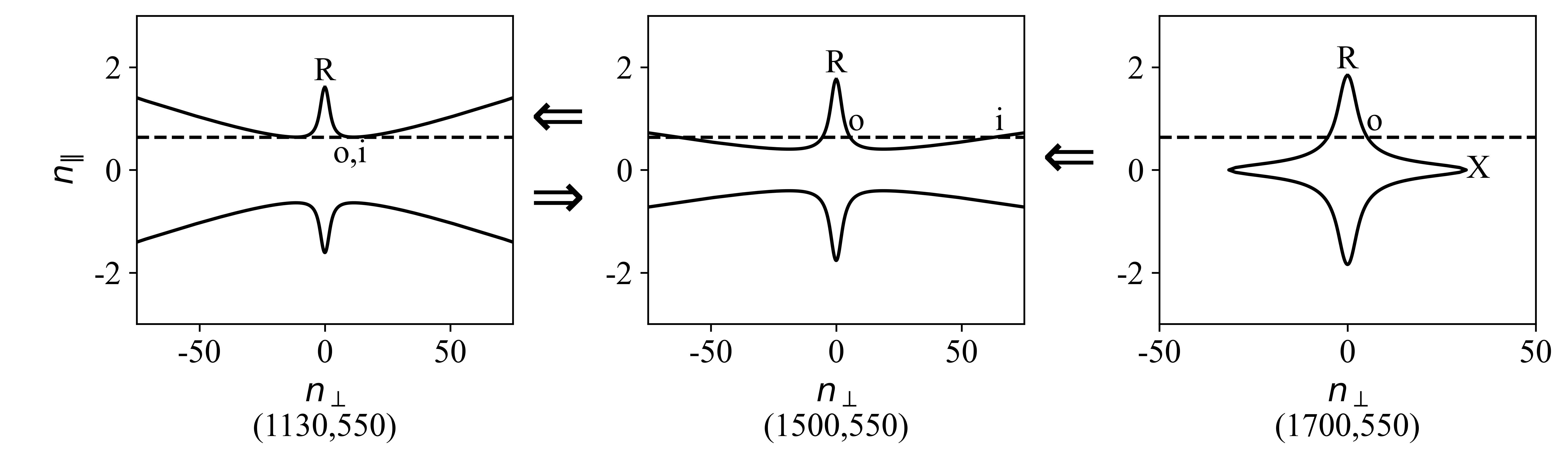}
    \end{minipage}
  }
  \subfigure[$n_{\|}=2.0$]
  {
    \begin{minipage}[b]{.45\textwidth}
      \centering
      \includegraphics[width=\textwidth]{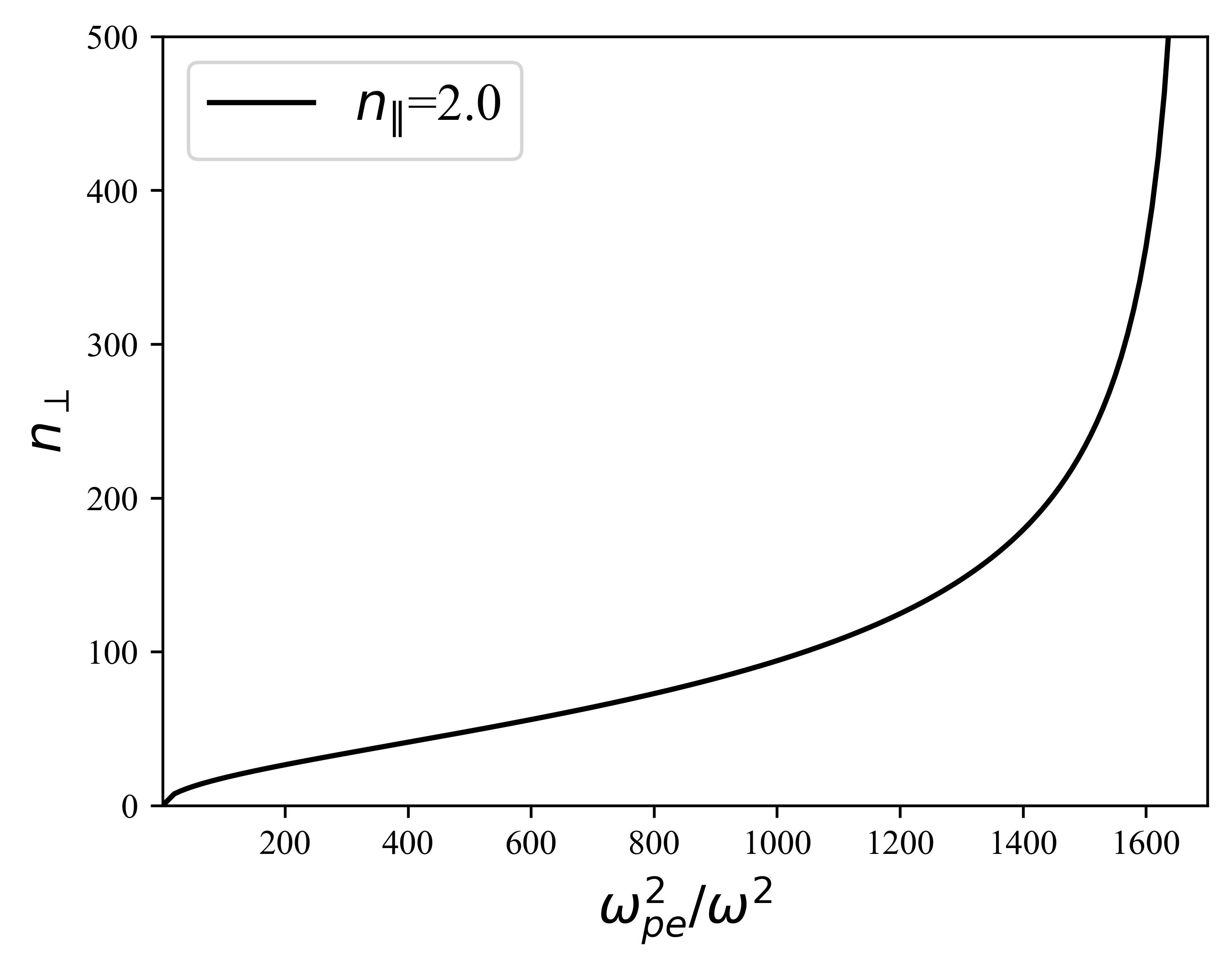}
    \end{minipage}
  }
  \subfigure[$n_{\|}=0.64$]
  {
    \begin{minipage}[b]{.45\textwidth}
      \centering
      \includegraphics[width=\textwidth]{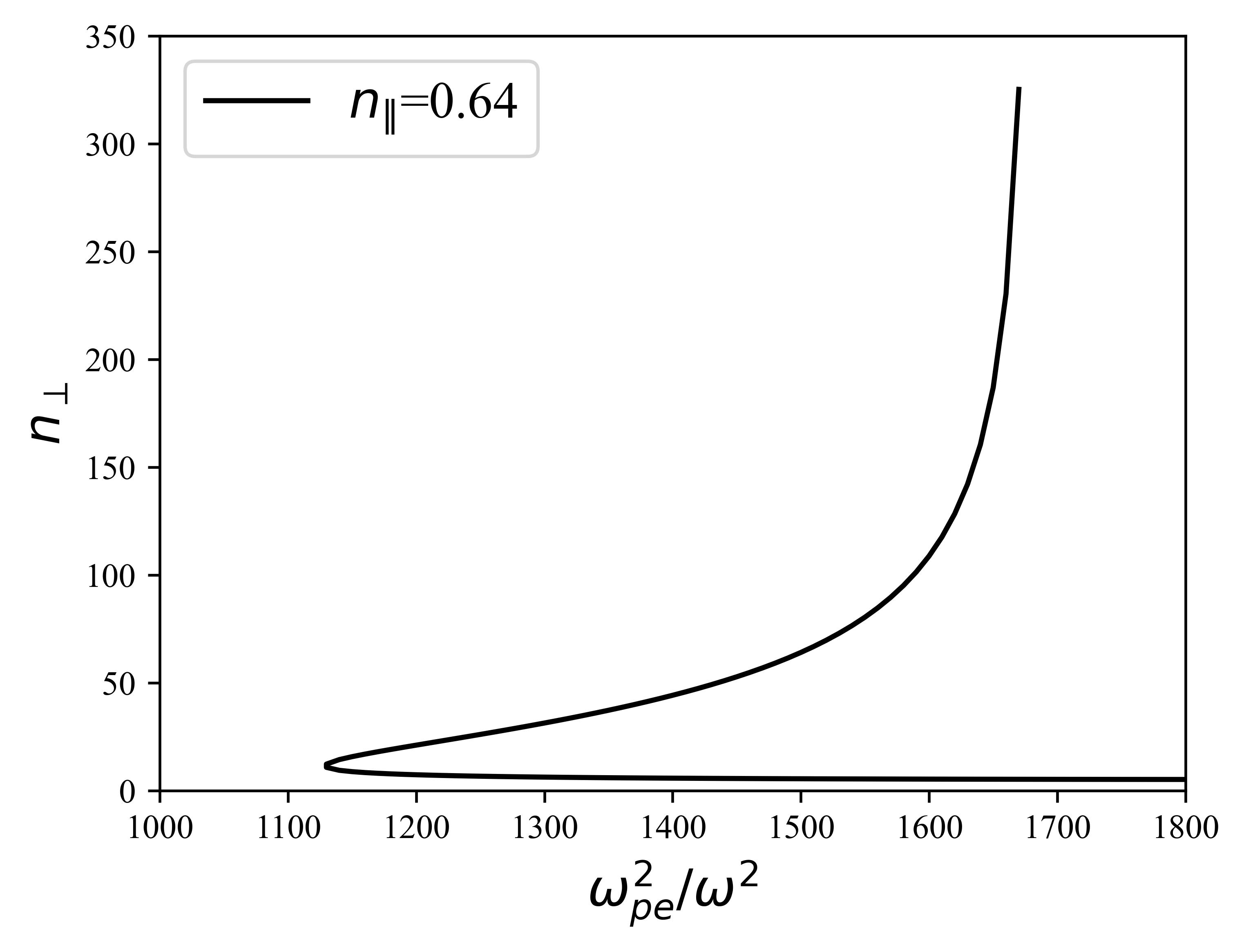}
    \end{minipage}
  }
  \caption{\label{fig:propagation paths 1} Figures (a) and (b) depict the IWNSs along the wave propagation paths in a cold plasma with $\Omega_{ce}/\omega=550$,$n_{\|}=2.0$ and $n_{\|}=0.64$. The coordinates below each subfigure indicate the position in the CMA diagram. In these figures, the symbol "o" represents waves propagating toward the lower density region (i.e., toward the outside of the plasma), while the symbol "i" indicates waves propagating toward the higher density region (i.e., toward the inside of the plasma). The arrows between each figure indicate the propagation direction, with a pair of forward and backward arrows representing wave reflection. Figures (c) and (d) display the $n_{\perp}$ versus $\omega_{pe}^{2}/\omega^{2}$ diagrams for the same waves represented in Figures (a) and (b).}
\end{figure*}

If the IWNS belongs to type 1b (2b) in Fig.\ref{fig:iwns class}, and the fixed $n_{\|}$ line is tangent to the IWNS at a $\theta\neq 0$ extreme point, the wave will be reflected and undergo a transition from a forward wave to a backward wave.
Specifically, Fig.\ref{fig:propagation paths 1} traces wave paths under density gradients, focusing on slow waves in regions (10) and (12) of Fig.\ref{fig:cold cma diagram}
As shown in the subfigure with coordinates $(1500,550)$ in Fig.\ref{fig:propagation paths 1} (b), the dashed line representing $n_{\|}=0.64$ intersects the R wave IWNS at two points in one quadrant of the $(n_{\perp},n_{\|})$ plane, denoted by `o' and `i'.
The intersection point with the larger $|n_{\perp}|$ represents the slow wave, while the point with the smaller $|n_{\perp}|$ corresponds to the fast wave.
The $n_{\perp}$ projections of the normal directions at points `o' and `i' are opposite.
Consequently, the two waves represented by `o' and `i' have opposite directions of group velocity-one propagating forward and the other backward.
When the dashed line for $n_{\|}=0.64$ is tangent to the R wave IWNS, points `o' and `i' coincide, signifying a mode transformation between the waves denoted by `o' and `i'.
This represents the forward-to-backward transition.
Therefore, the wave propagation story of Fig.\ref{fig:propagation paths 1} (b) can be summarized as follows: 

(1)A wave (point `o') with $n_{\|}=0.64$ is generated in the regime of $\omega_{pe}^{2}/\omega^{2}=1700$ and $\Omega_{ce}/\omega=550$, after which it propagates into a lower density regime.

(2)When the wave reaches the regime of $\omega_{pe}^{2}/\omega^{2}=1130$ and $\Omega_{ce}/\omega=550$, it is reflected  (point `o' $\rightarrow$ `i').

(3)The wave then propagates toward a higher density regime but cannot cross the lower hybrid resonant line.

In general, the slow wave propagation depends on whether $n_{\|}$ is above or below a certain threshold.
In regions (10) and (12) of Fig.\ref{fig:cold cma diagram}, the minimum value $|n_{\|}|_{min}$ on the R wave IWNS can be derived from Eq.(\ref{eq:hook function 1})
\begin{equation}
  |n_{\|}|_{min}=\sqrt{\frac{S(P-R)(P-L)}{(P-S)^2}}+\sqrt{\frac{-PD^{2}}{(P-S)^2}}.
  \label{eq:n_min}
\end{equation}
If $|n_{\|}|$ exceeds $max\{|n_{\|}|_{min}\}$ in Eq.(\ref{eq:n_min}), the slow wave is able to propagate from $\omega_{pe}^{2}/\omega^{2}=1$ to the lower hybrid resonant line, as depicted by Fig.\ref{fig:propagation paths 1} (a) and (c).
However, the slow wave cannot cross the lower hybrid resonant line due to the R wave IWNS transition from type 2b to type 1c, and divergence of $n_{\perp}^{2}$.
This can be seen from Fig.2 and Fig.3, where the slow wave $n_{\perp}^{2}$ approaches infinity on the lower hybrid resonant line.
On the other hand, if $|n_{\|}|$ satisfies $|n_{\|}|<max\{|n_{\|}|_{min}\}$ in Eq.(\ref{eq:n_min}), the constant $n_{\|}$ line can be tangent to the R wave IWNS at the minimum point. 
In this case, the outward fast wave will convert into the inward slow wave as depicted by Fig.\ref{fig:propagation paths 1} (b) and (d).

\section{The CMA Diagram of Thermal Plasma}
The dielectric tensor for a single ion species plasma at temperature $T$, as described in references \cite{stix1992,swanson}, is employed.
To solve the dispersion relation, we use the generalized argument principle code ZPL (Zero-Pole-Location) \cite{chenhaotian2022}.
This code has the ability to locate all the zeros and poles of a meromorphic function within a closed complex domain.
In this work, we consider an isotropic $\left(T_{\perp}=T_{\|}\right)$ homogeneous plasma, which is composed of electrons and hydrogen ions, with the mass ratio ${M}/{m}=1836$ and the temperature ratio ${T_{i}}/{T_{e}} = 1$.

\subsection{Motivation}
\begin{figure}
  \centering
  \subfigure[$n_{\|}=2.0,\Omega_{ce}/\omega=550$]
  {
    \begin{minipage}[b]{0.45\textwidth}
      \centering
      \includegraphics[width=\textwidth]{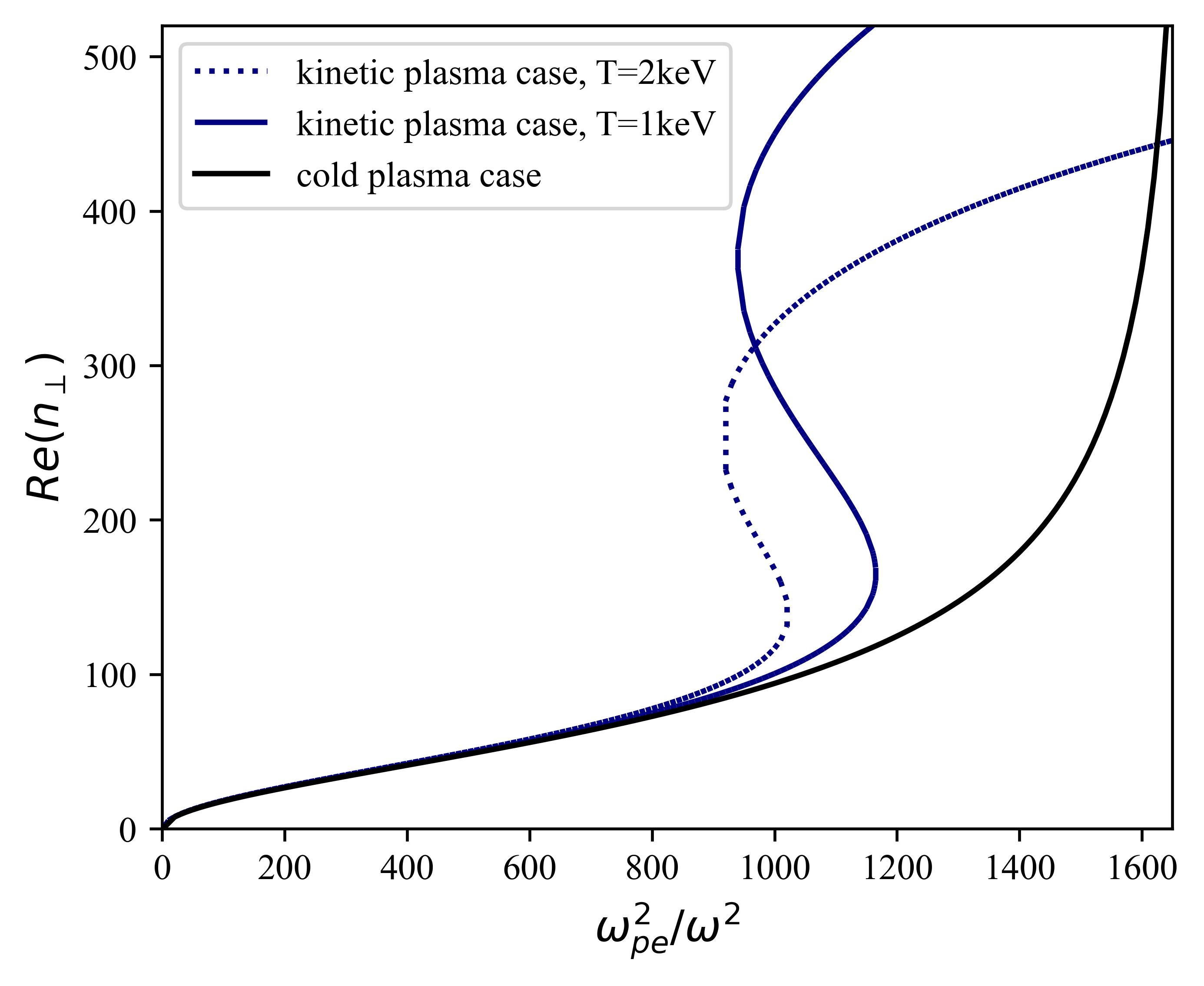}
    \end{minipage}
  }
  \subfigure[$n_{\|}=3.0,\Omega_{ce}/\omega=1.2$]
  {
    \begin{minipage}[b]{0.48\textwidth}
      \centering
      \includegraphics[width=\textwidth]{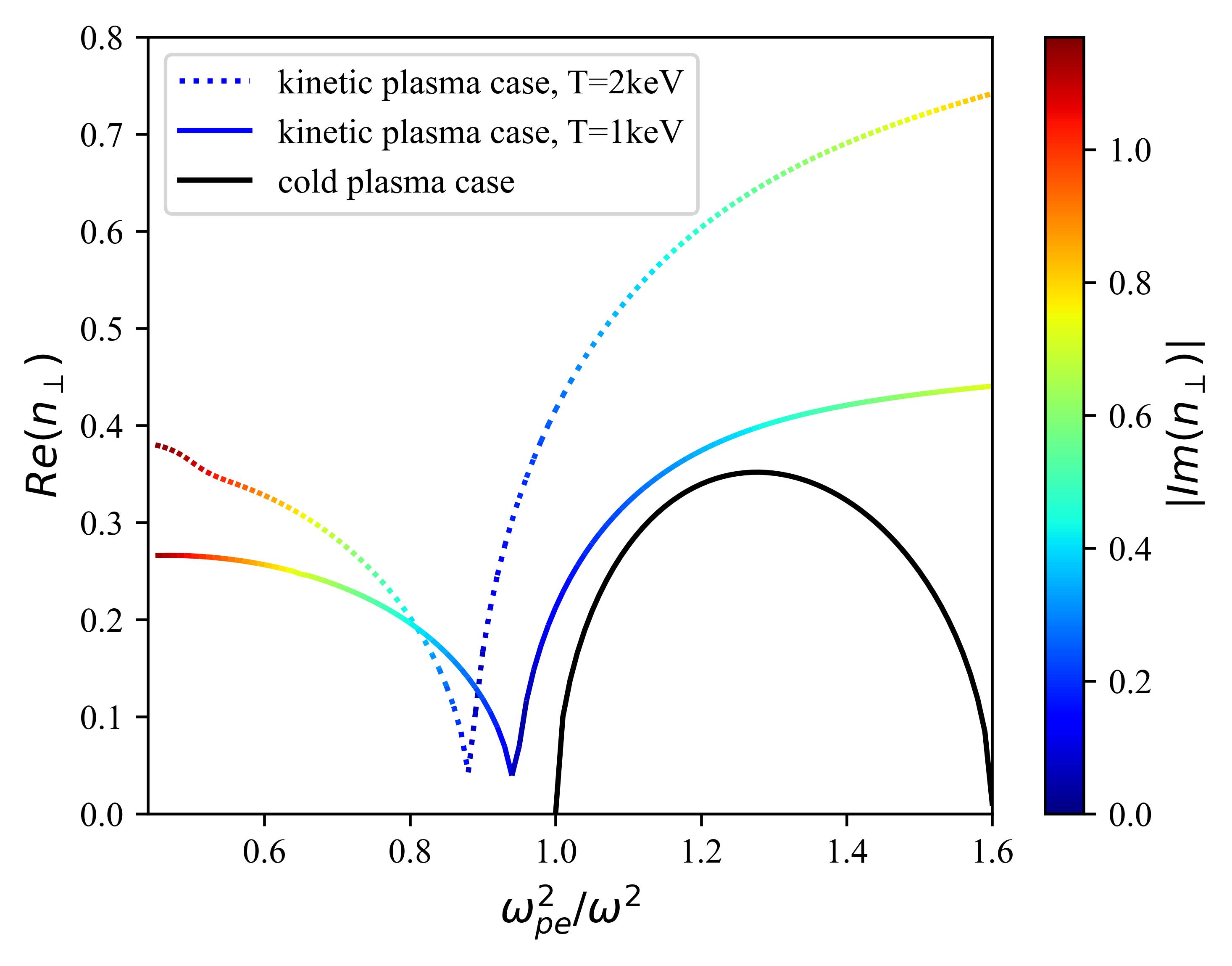}
    \end{minipage}
  }
  \caption{\label{motivation} Two comparisons of wave propagation paths under density gradients between kinetic and cold plasma cases. Case 1 demonstrates wave propagation paths for $n_{\|}=2.0,\Omega_{ce}/\omega=550$. Meanwhile, Case 2 represents wave propagation paths with $n_{\|}=3.0,\Omega_{ce}/\omega=1.2$.}
\end{figure}

The original CMA diagram does not incorporate kinetic effects,and it is thus inapplicable for analysing the wave propagation of kinetic waves.
To clarify this problem, we present two comparisons of wave propagation paths under density gradients between kinetic and cold plasma cases, as depicted in Fig.\ref{motivation}.

In Fig.\ref{motivation} (a), we illustrate the propagation paths of the slow wave at multiples of the ion cyclotron frequency.
In the cold plasma case, as discussed in the previous section, the slow wave can propagate from $\omega_{pe}^{2}/\omega^{2}=1$ to the lower hybrid resonant line. However, it cannot cross this line.
When kinetic effects are taken into account, the slow wave can experience two mode transformations into the ion Bernstein wave. 
Different temperatures only lead to quantitative changes rather than introducing qualitative changes on the wave propagation paths.
Unlike the slow wave in the cold plasma case, the ion Bernstein wave has no limitation at the lower hybrid resonant line and can propagate to regions of higher density.
This essential difference implies that the finite Larmor radius (FLR) effect should be considered in the CMA diagram.

Furthermore, the original CMA diagram does not account for the cyclotron damping effect. 
In Fig.\ref{motivation} (b), we show the propagation paths of a quasi-parallel wave with a frequency close to the electron cyclotron frequency.
In the cold plasma scenario, this wave exists in the regime where $\omega_{pe}^{2}/\omega^{2}>1$ and propagates between two cutoff points (where $n_{\perp}=0$). 
In contrast, when considering the kinetic effects, the wave experiences damping in the $\omega_{pe}^{2}/\omega^{2}<1$ regime. 
As the wave propagates to the $\omega_{pe}^{2}/\omega^{2}\simeq 1$ regime, it can transform into the Langmuir wave, and simultaneously the damping rate decreases.
After crossing the $\omega_{pe}^{2}/\omega^{2}=1$ boundary and continuing to propagate into the higher density regime, the damping rate increases. 
These phenomena illustrate the characteristics of the energy deposition region.
To comprehensively summarize these properties of kinetic waves and provide a more effective framework for analysing wave propagation, it is thus necessary to develop a kinetic CMA diagram.

\subsection{A Summary of Kinetic Effects}

\begin{figure}
  \centering
  \includegraphics[width=0.5\textwidth]{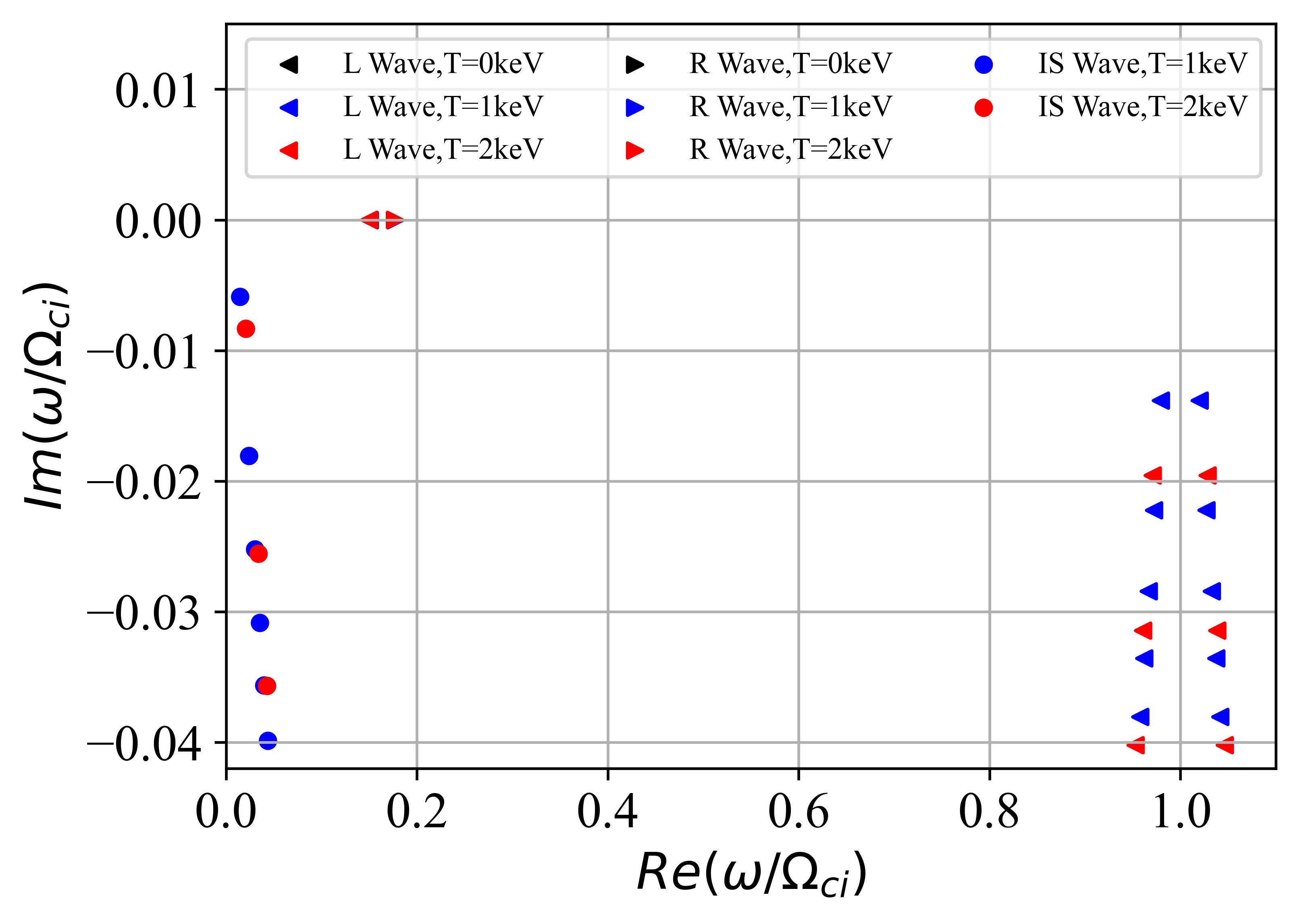}
  \caption{\label{fig:kinetic_parallel_wave_1}  The parallel wave frequency distribution for $k_{\|}c/\Omega_{ci}=7,{\omega^{2}_{pe}}/{\Omega^{2}_{ce}}=1$ in the region of $0.001<{\omega_r}/{\Omega_{ci}}<1.2,-0.04<{\omega_i}/{\Omega_{ci}}<0.01$. Here, the IS symbol represents the ion-sound wave. 
  Waves at different temperatures are denoted by different colours. 
  In the case of $T=0 $keV (cold plasma case), the normal modes are located on the real axis.
  Notably, these normal modes are close to the shear alfven wave frequency ($\omega=k_{\|}v_{A}$) and are nearly identical to the results obtained for $T=1$keV and $T=2 $keV.}
\end{figure}

Figure \ref{fig:kinetic_parallel_wave_1} depicts a typical distribution of kinetic wave frequencies in the complex plane, with the parallel kinetic waves as a representative example.
When considering a finite temperature condition, the ion sound wave branch emerges.
Notably, the specific characteristics of the ion sound wave branch are directly dependent on the temperature. 
The heavily damped modes, particularly those situated beside the $\omega/\Omega_{ci}=0$ and $1$ axes, are strongly influenced by the wave particle resonance. 
The presence of an infinite number of modes is a prevalent and well-recognized phenomenon within the kinetic dispersion relation \cite{Chenhaotian2021}. This fact emphasizes the complexity inherent in the study of kinetic waves and their associated dispersion characteristics.
In order to streamline the analysis and maintain a clear focus, we make a deliberate choice to exclude these heavily damped modes from our immediate consideration. Instead, within the framework of the CMA diagram for thermal plasma, our attention is concentrated solely on the least damped modes. 

1.Kinetic effects on parallel waves

\begin{figure}
    \includegraphics[height=0.3\textheight]{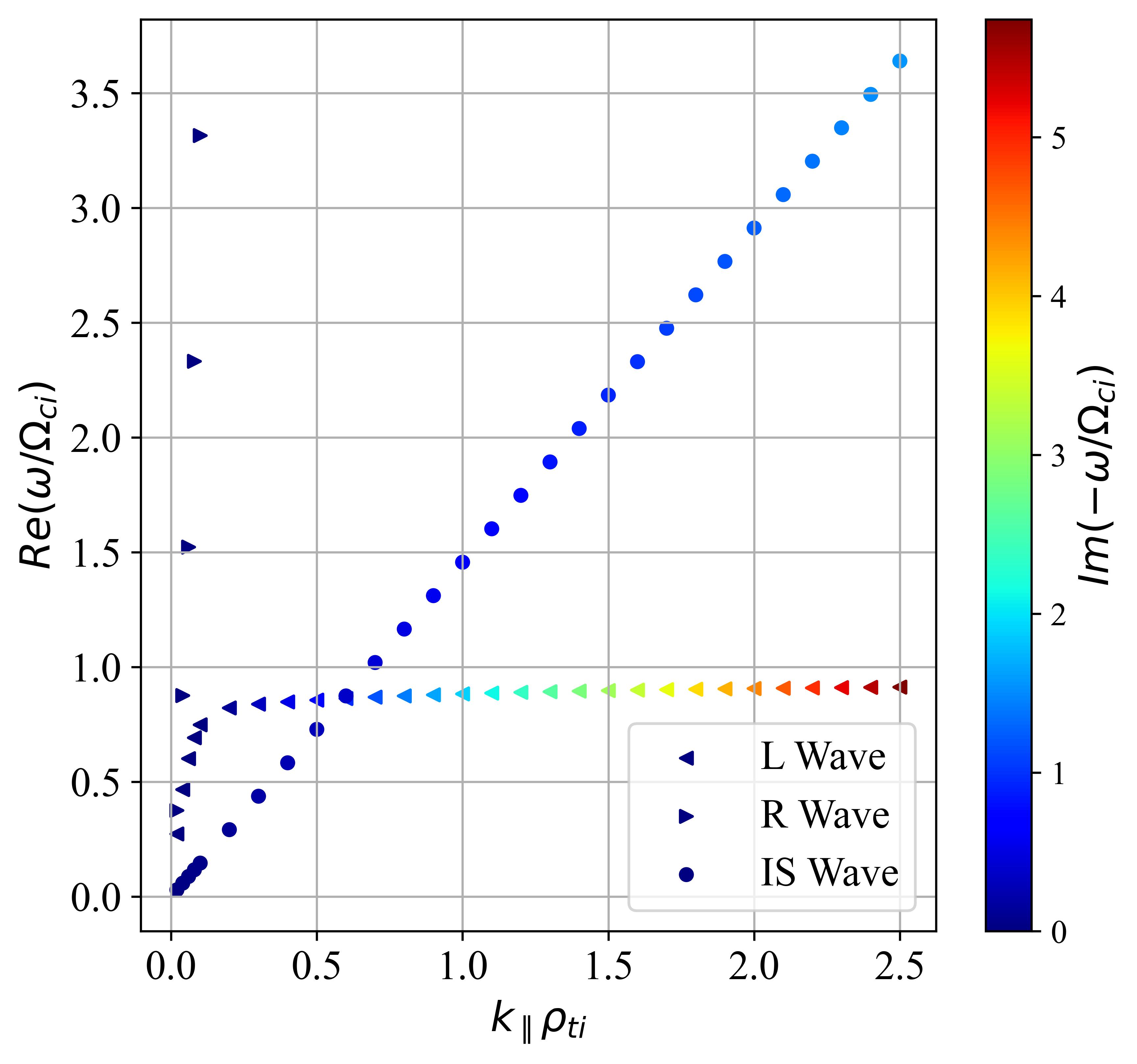}
    \includegraphics[height=0.3\textheight]{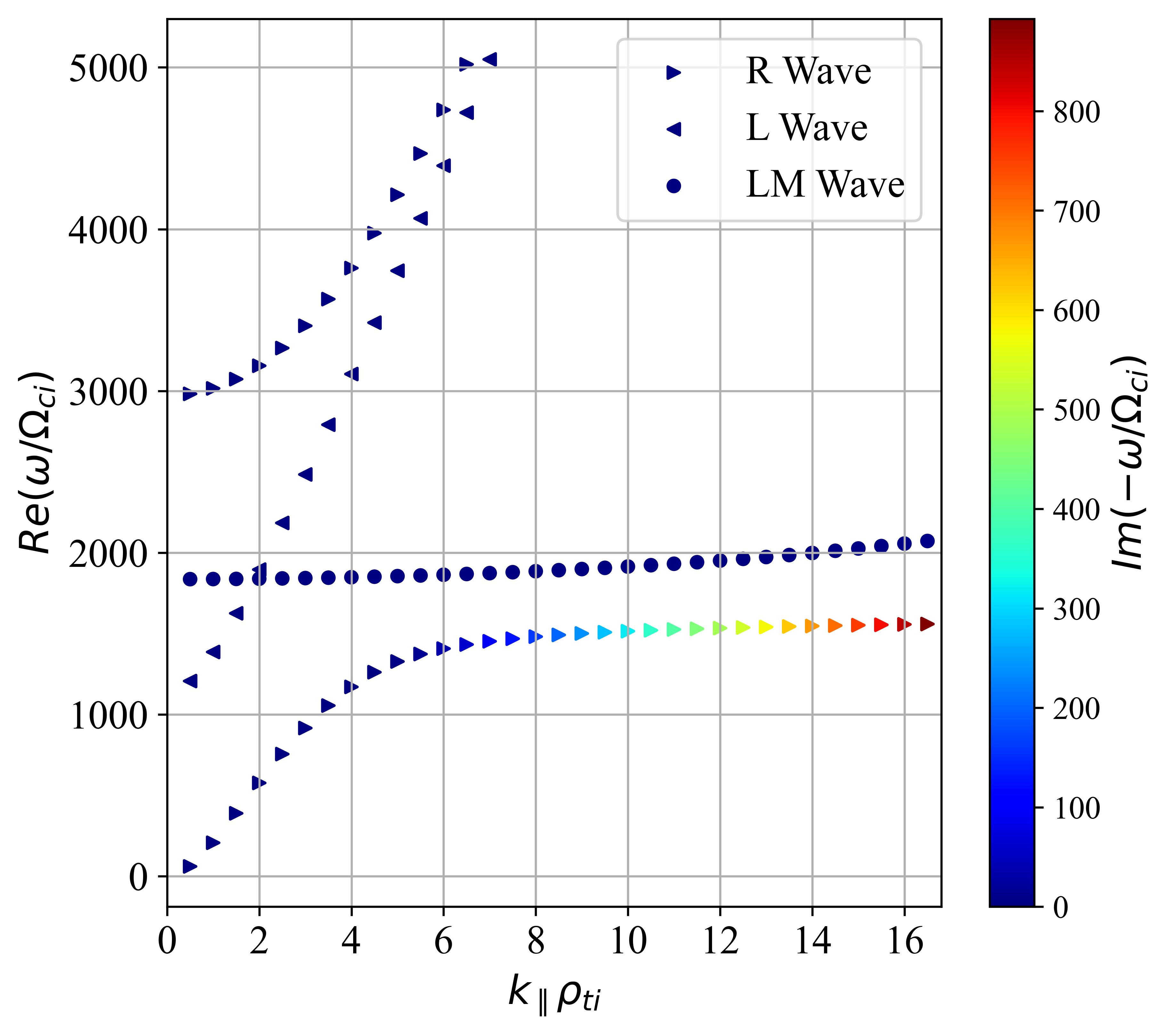}
    \caption{\label{fig:kinetic_parallel_wave_2}  The least damped L, R, Ion-sound (IS) and Langmuir (LM) wave frequencies versus $k_{\|}\rho_{ti}$ diagrams for ${\omega^{2}_{pe}}/{\Omega^{2}_{ce}}=1,T_{i}=1keV$ in both low and high frequency regimes. 
    In this diagrams, the IS and LM symbols denote the ion-sound and Langmuir waves, respectively. The color of each point represents the damping rate $Im(-{\omega}/{\Omega_{ci}})$.}
\end{figure}

Figure \ref{fig:kinetic_parallel_wave_2} presents the kinetic effects on the parallel waves. 
For the low-frequency L (R) wave branch, the damping rate increases as the real frequency nears the ion (electron) cyclotron frequency because of cyclotron damping. 
When the real frequency moves away from the cyclotron frequency, these low-frequency L and R waves exhibit similar behaviour as cold plasma waves. 
Therefore, the kinetic effect is significant when the real frequency is close to the cyclotron frequency.
On the other hand, the high-frequency L and R waves are not damped since the phase velocities exceed the speed of light, and no particles can resonate with the waves.
Unlike in the cold plasma case, both the damping rate and the real frequency of the Langmuir wave branch increase with $k_{\|}\rho_{ti}$ due to Landau damping.

2.Kinetic effects on perpendicular waves

\begin{figure}
  \includegraphics[height=0.3\textheight]{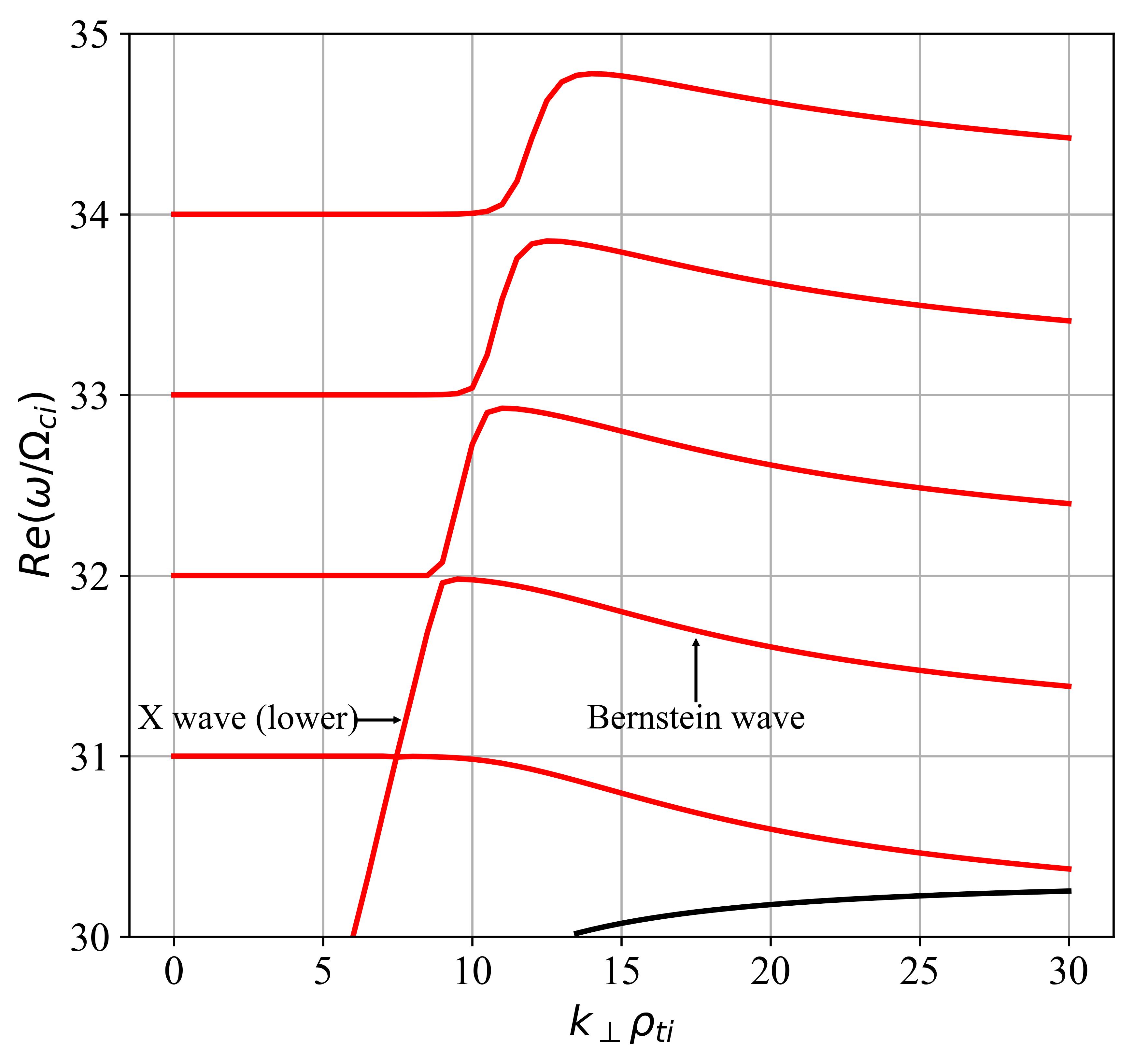}
  \includegraphics[height=0.3\textheight]{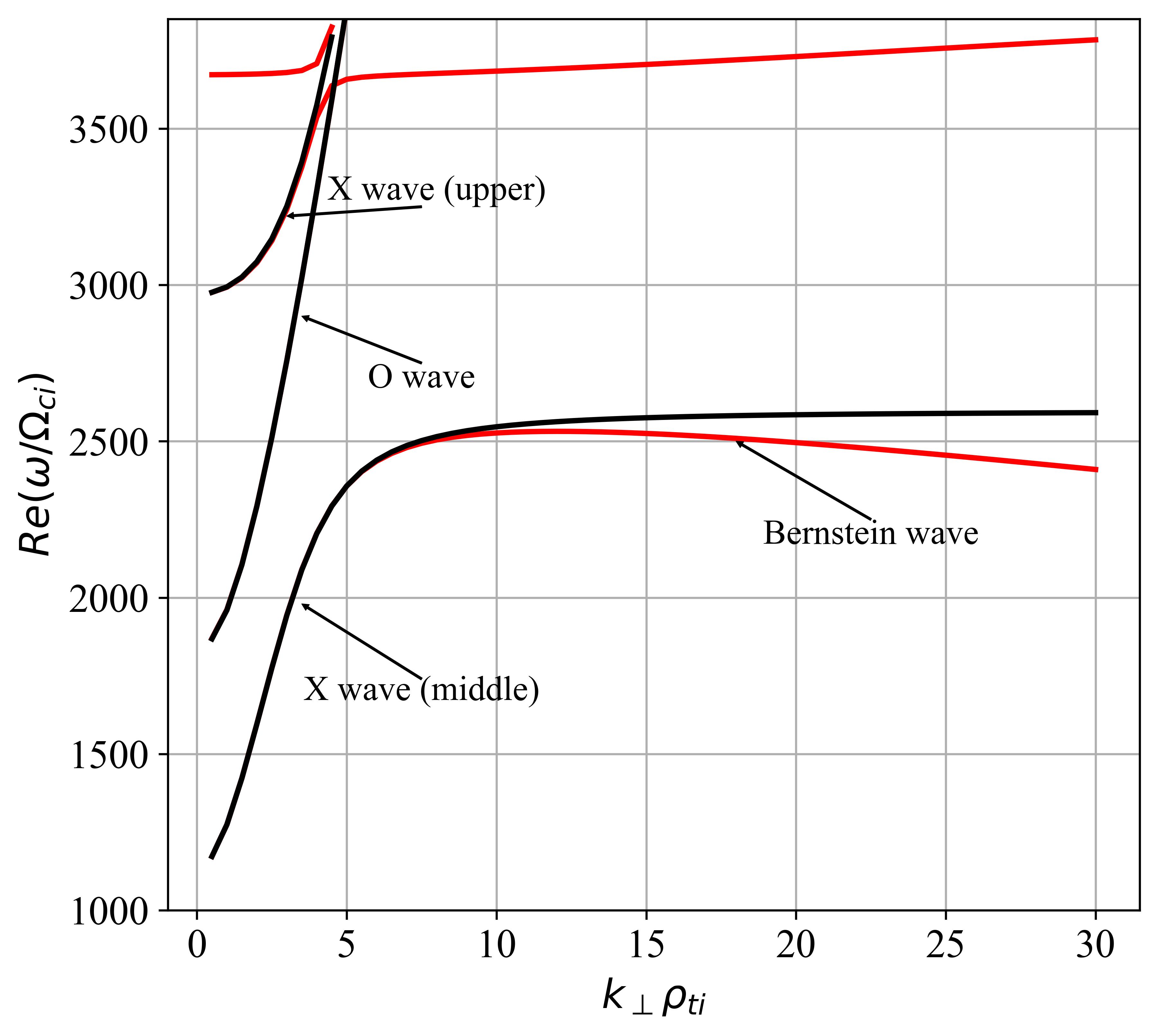}
  \caption{\label{fig:kinetic_perpendicular_wave} The perpendicular wave frequencies versus $k_{\perp}\rho_{ti}$ diagram for ${\omega^{2}_{pe}}/{\Omega^{2}_{ce}}=1,T_i=1keV$ in both low and high frequency regimes. The red curves denote the solutions of the kinetic dispersion relation while the black curves denote the cold plasma waves.}
\end{figure}

Figure \ref{fig:kinetic_perpendicular_wave} displays the kinetic effects on the perpendicular waves.
In the proximity of the lower and upper hybrid frequencies, the polarization of the  X wave branch tends toward ${E_{1y}}/{E_{1x}}<<1,{E_{1z}}/{E_{1x}}=0$, and the kinetic X wave ultimately transforms into a Bernstein wave due to the FLR effect.
The properties of Bernstein waves are different depending on whether they are above or below the extraordinary-Bernstein (X-B) mode transformation frequency [6]. 
For the Bernstein wave above the X-B mode transformation frequency, its frequency first increases from $\omega=n\Omega_{ci}$ line and then decreases back to $\omega=n\Omega_{ci}$ line.
In contrast, the Bernstein wave below the X-B mode transformation frequency decreases from $\omega=n\Omega_{ci}$ to $\omega=(n-1)\Omega_{ci}$.
As a result, the boundaries of the X wave are no longer defined by the hybrid resonant frequency lines in the CMA diagram of thermal plasmas. 
Instead, they are determined by the X-B mode transformation frequency lines.

3.Kinetic effects on oblique propagation waves

\begin{figure}
    \centering
    \subfigure[Different temperatures]
    {
    \begin{minipage}[b]{0.47\textwidth}
      \centering
      \includegraphics[width=\textwidth]{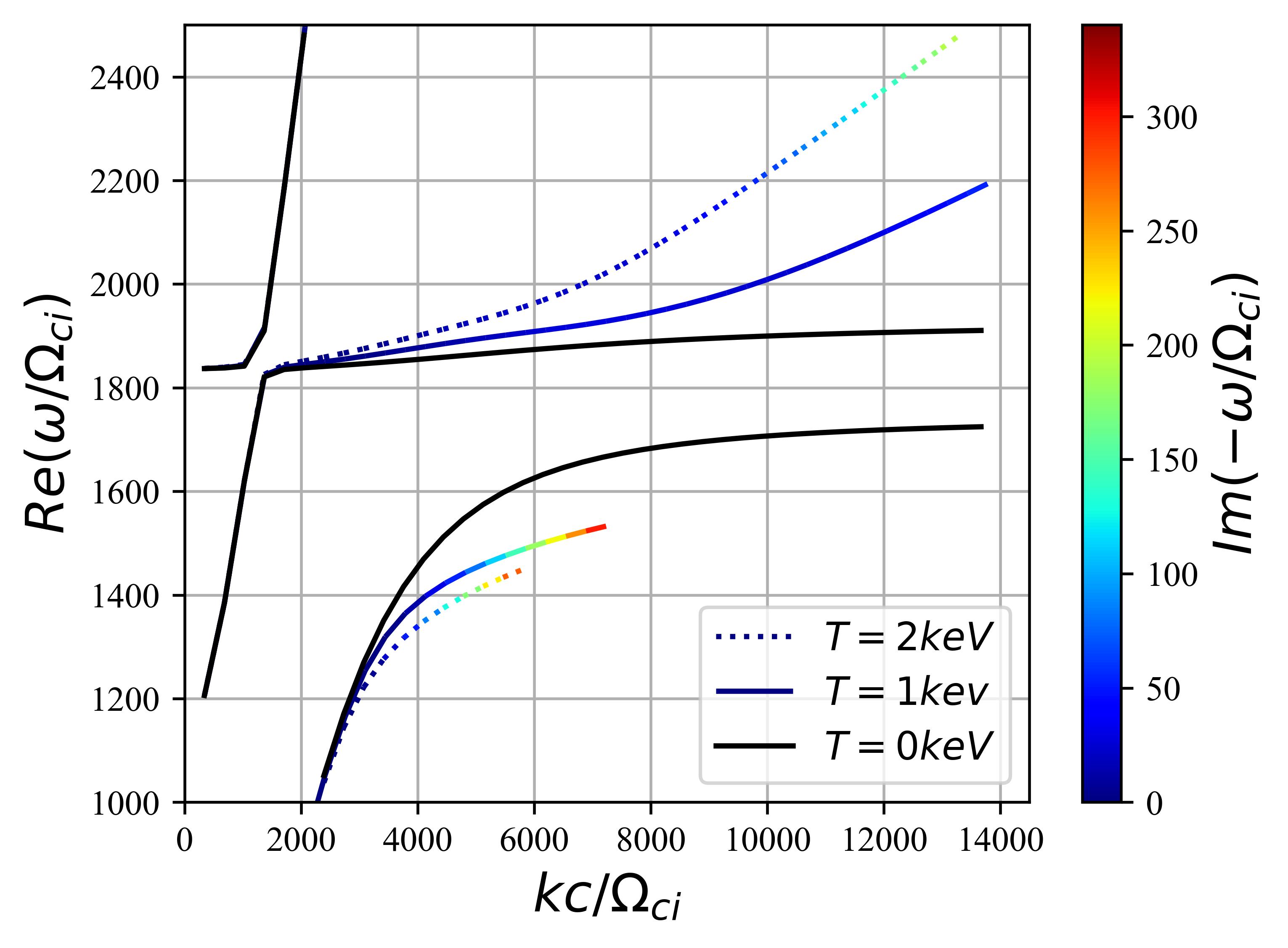}
    \end{minipage}
    }
    \subfigure[Different propagation angles]
    {
    \begin{minipage}[b]{0.47\textwidth}
      \centering
      \includegraphics[width=\textwidth]{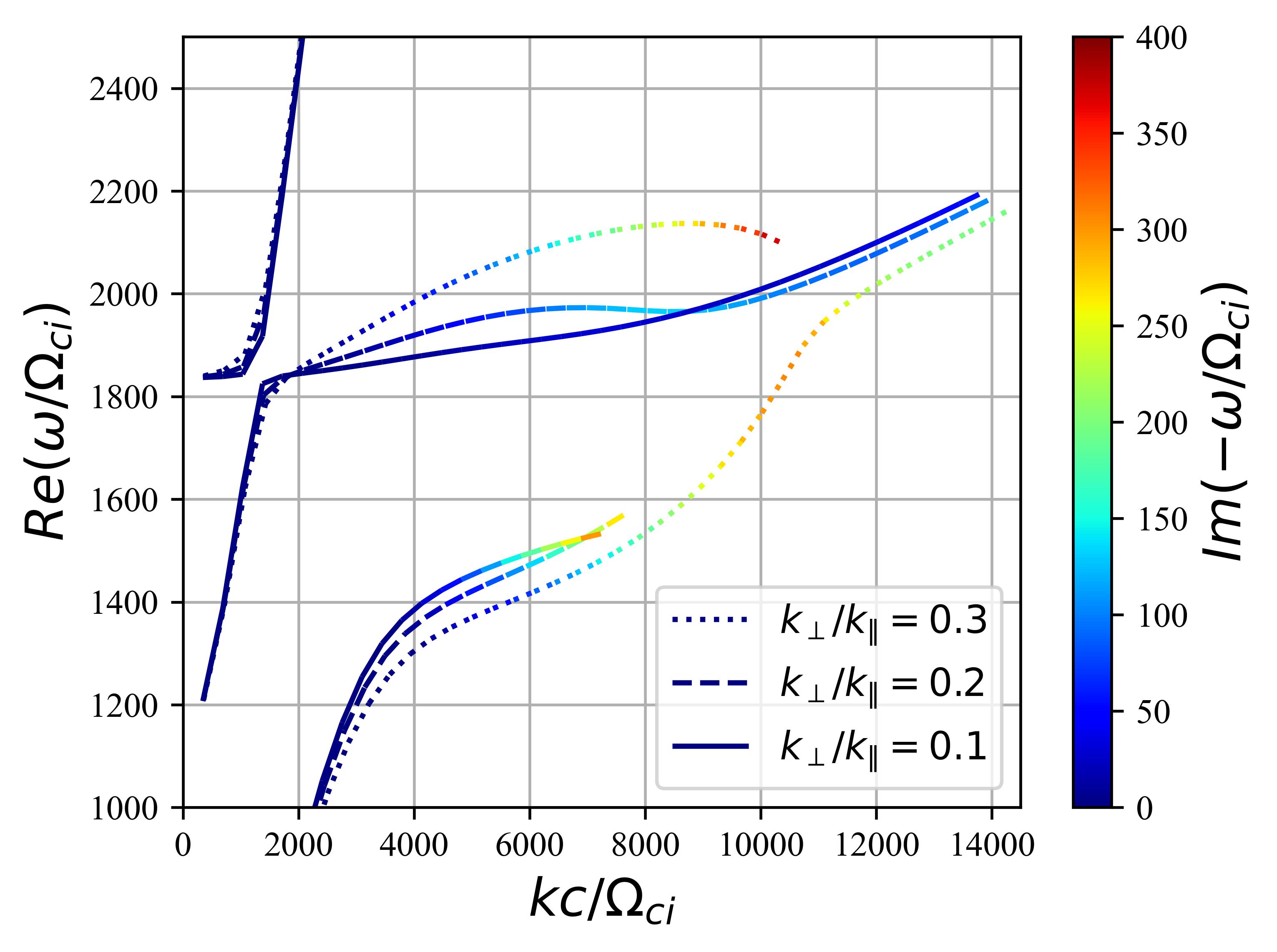}
    \end{minipage}
    }
    \caption{\label{fig:kinetic general wave}  The effects of different temperatures and propagation angles on oblique propagation waves. 
    Figure (a) demonstrates the impact of temperature on the quasi-parallel wave where $k_{\perp}/k_{\|}=0.1$ and ${\omega^{2}_{pe}}/{\Omega^{2}_{ce}}=1$. 
    Figure (b) shows the effect of the propagation angle effect on the kinetic wave when ${\omega^{2}_{pe}}/{\Omega^{2}_{ce}}=1$ and $T=1keV$. 
    The ZPL code search domain is $1000<Re({\omega}/{\Omega_{ci}})<2500,-30k_{\|}\rho_{ti}<Im({\omega}/{\Omega_{ci}})<0$.}
\end{figure}

Figure \ref{fig:kinetic general wave} depicts the impacts of temperature and propagation angles on oblique propagation waves, using the quasi-parallel waves as an illustrative example.

(1)Effect of temperature

The magnitude of the temperature is a direct indicator of the strength of kinetic effects. 
As can be seen in Fig.\ref{fig:kinetic general wave} (a), when the temperature increases, both the Landau damping and cyclotron damping effects become more pronounced.
This increase in damping effects has a significant impact on the propagation characteristics and energy dissipation rates.

(2)Effect of propagation angles

When $k_{\perp}/k_{\|}$ increases from 0, the curves of the Langmuir wave and the L wave split into two distinct segments, as shown in Fig.\ref{fig:kinetic general wave} (b).
In cold plasmas, as $k_{\perp}/k_{\|}\rightarrow +\infty$ , the first segment of the Langmuir wave, together with the second segment of the L wave, undergoes a transformation into the O wave. 
Simultaneously, the second segment of the Langmuir wave, combined with the first segment of the L wave, becomes the X wave. 
This transformation is a characteristic feature of wave behaviour in cold plasmas.
Similarly, in thermal plasmas, when the frequency is close to the plasma frequency and ${k_{\perp}}/{k_{\|}}\rightarrow +\infty$, the Langmuir wave can smoothly transform into either the O wave or the X wave. 

A different phenomenon occurs when the frequency is far from the plasma frequency. 
As ${k_{\perp}}/{k_{\|}}$ increases, the Langmuir wave curve experiences substantial damping. 
It then descends and merges with the R wave curve. 
In this case, when the frequency is far from the plasma frequency, the Langmuir wave can't transform continuously into the O wave and X wave.

Similarly, by analysing the dispersion curves of quasi-perpendicular waves, we can see that when the ratio $k_{\|}/k_{\perp}$ is larger, the X-B mode transformation phenomenon no longer occurs.
In this case, as described in the reference \cite{Bernstein}, the Bernstein wave behaves as a quasi-perpendicular wave.

\subsection{Constructing The Kinetic CMA Diagram}

\begin{figure}
  \centering
  \subfigure[Overall figure]
  {
    \begin{minipage}[b]{0.6\textwidth}
      \centering
      \includegraphics[width=\textwidth]{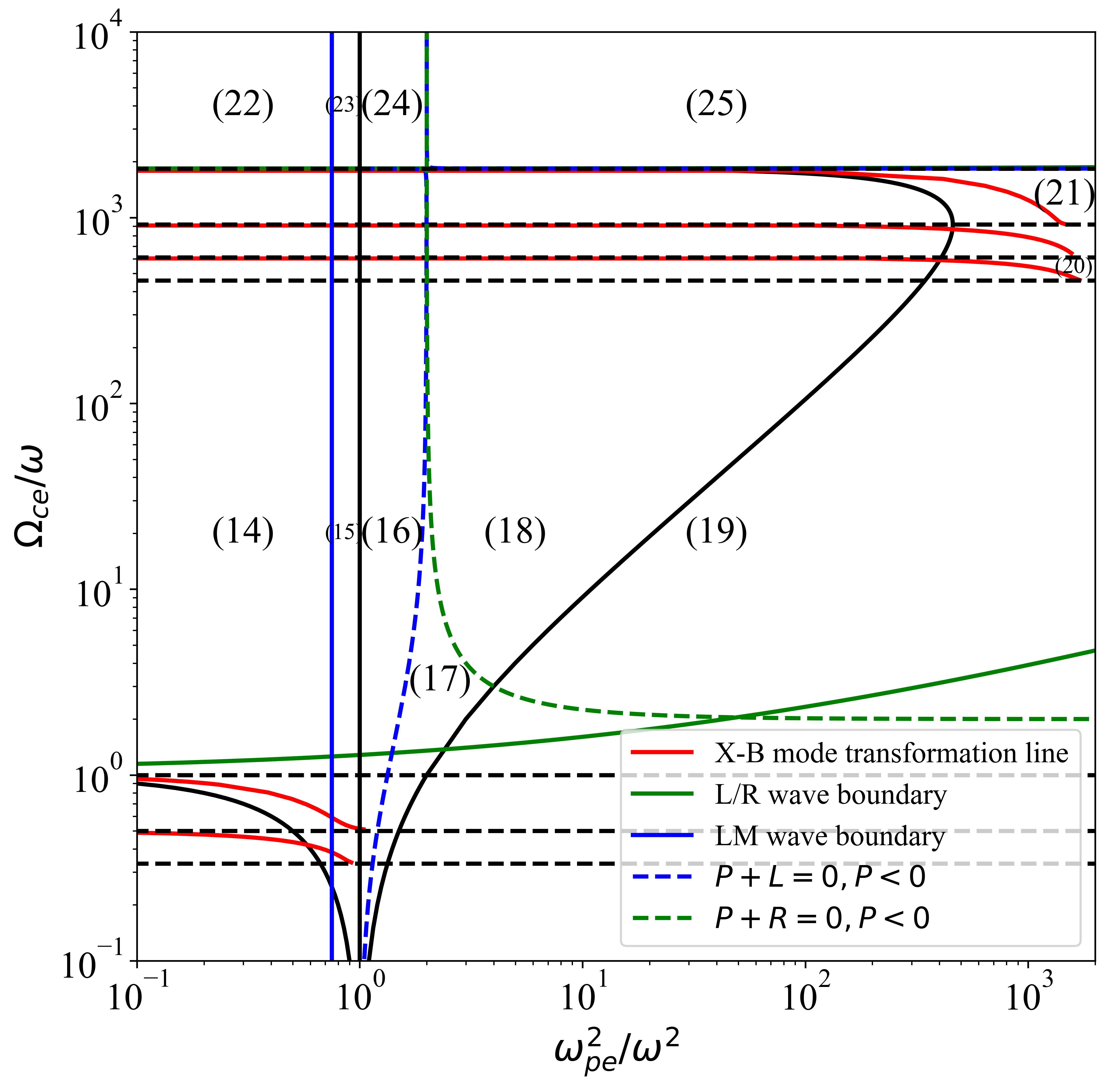}
    \end{minipage}
  }
  \subfigure[Enlarged view of the high-frequency regime]
  {
    \begin{minipage}[b]{0.6\textwidth}
      \centering
      \includegraphics[width=\textwidth]{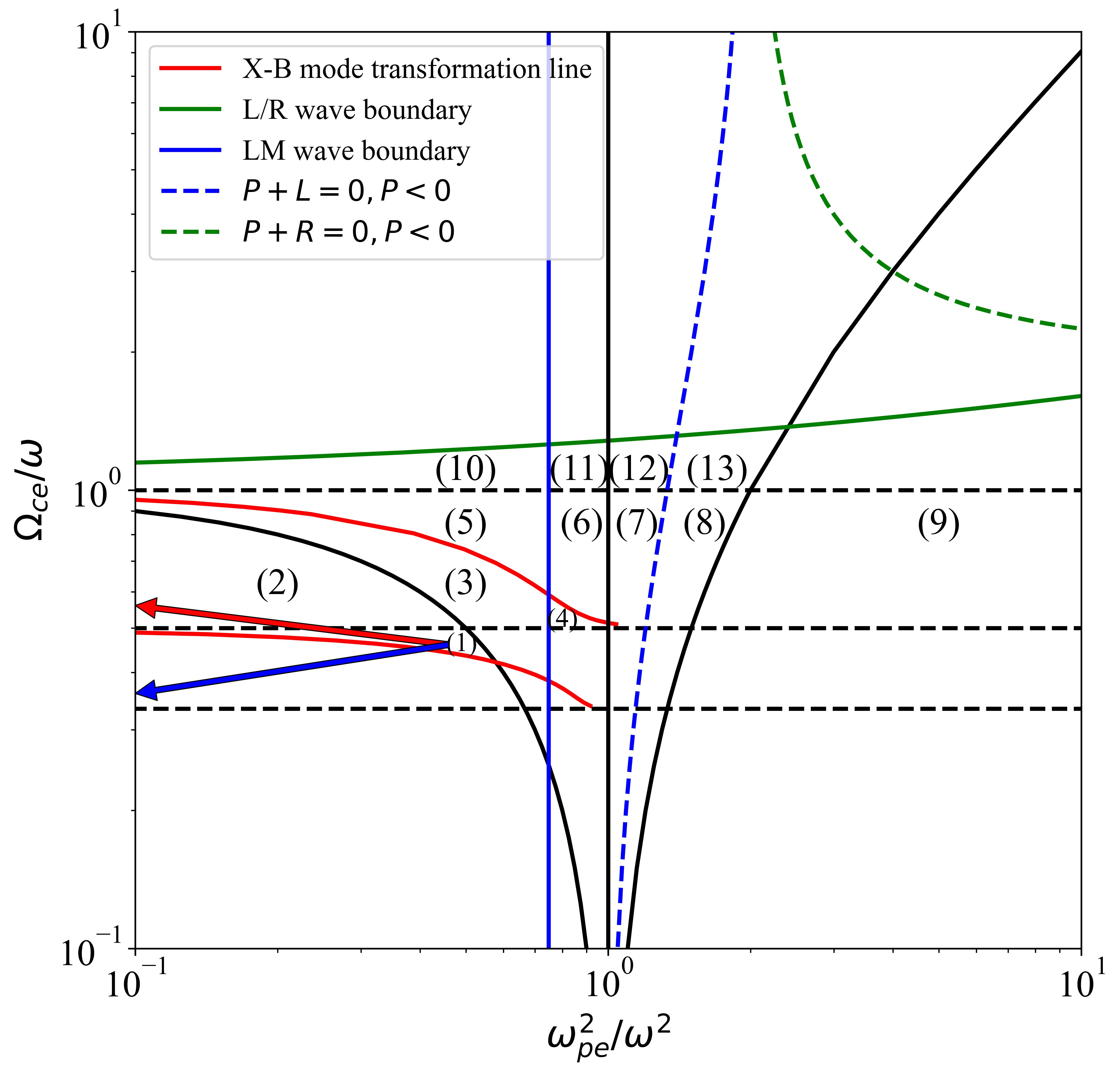}
    \end{minipage}
  }
  \caption{\label{fig:kinetic cma diagram}  The kinetic CMA diagram with new L, R and LM wave boundaries and X-B mode transformation lines for both high frequency and low frequency regimes. These diagrams only retain the first few mode transformation lines. In the second subfigure, the red and blue arrows represent the high field side path and the low field side path, respectively. These paths are discussed in the following section. It should be noted that the regions (1-25) presented in Fig.\ref{fig:cold cma diagram} and Fig.\ref{fig:kinetic cma diagram} do not have a one-to-one correspondence.}
\end{figure}

The kinetic dispersion relation brings in an extra parameter, temperature $T$. 
Although the value of $T$ has a quantitative impact on the CMA diagram, the topological features of the boundaries and IWNSs stay the same regardless of different temperatures, as discussed previously. 
Thus, to clarify the fundamental characteristics of the kinetic CMA diagram, we consider the $T=1keV$ case.

Noting that the kinetic dispersion relation converges to the cold plasma dispersion relation as $k \rightarrow 0$, the cutoff frequencies in the kinetic CMA diagram are the same as those in the original CMA diagram. 
On the contrary, kinetic effects can influence the resonant frequencies.

\paragraph{Parallel Waves}

We retain weakly damped waves that meet the artificially chosen criterion ${\omega_{i}}/{\omega_{r}}>-0.01$. 
In the kinetic CMA diagram, the new boundaries of the L, R, and Langmuir waves can't be determined analytically. 
Therefore, we use the following numerical calculation strategy:

First, we set ${\omega_{pe}^{2}}/{\Omega_{ce}^{2}}$ as a fixed value and gradually increase $k_{\|}\rho_{ti}$ from 0. 
Since the damping rates of the L, R, and Langmuir waves increase with $k_{\|}\rho_{ti}$, we can use the ZPL code to find the frequencies of these waves that satisfy $\omega_{i}/\omega_{r}=-0.01$.
This gives us one point $(\omega_{pe}^{2}/\omega^{2},\Omega_{ce}/\omega)$ on each new boundary.
Then, we scan different values of ${\omega_{pe}^{2}}/{\Omega_{ce}^{2}}$ and repeat the first step. This forms a nested loop that continues until we calculate the entire boundaries.

The new resultant boundaries of the L, R, and Langmuir waves are shown in Fig.\ref{fig:kinetic cma diagram}. 
Due to the cyclotron damping, the new boundary of the L (R) wave in the kinetic CMA diagram lies above the ion (electron) cyclotron frequency line, which is represented by the green curve in the Fig.\ref{fig:kinetic cma diagram}. 
The boundary of the Langmuir wave is independent of the magnetic field and appears as a vertical line (the yellow line in Fig.\ref{fig:kinetic cma diagram}). 
The weakly damped Langmuir wave exists between this vertical line and the $\omega^{2}_{pe}/\omega^{2}=1$ line.

\paragraph{Perpendicular Waves}

In the kinetic CMA diagram, the hybrid resonant frequency lines are replaced by the X-B mode transformation lines, shown as red curves in Fig.\ref{fig:kinetic cma diagram}. 
Additionally, the Bernstein wave is introduced as a quasi-perpendicular wave into the kinetic CMA diagram.
In principle, the X-B mode transformation frequencies are between $\omega=n\Omega_{ci}$ and $\omega=(n+1)\Omega_{ci}$ ($n=1,2,...$).
In Fig.\ref{fig:kinetic cma diagram}, only a few low-order mode transformation frequencies are shown as examples.
The domains of ion and electron Bernstein wave propagation are bounded by $\omega=n\Omega_{c}$, $\omega=(n+1)\Omega_{c}$ and the X-B mode transformation lines, i.e., the regions (5)+(6)+(7)+(8)+(9) in Fig.\ref{fig:kinetic cma diagram}.

Analytical expressions are no longer available for the boundaries of different IWNSs.
We choose to ignore the boundaries between type 1a and 1b*, and type 3a and 3b in Fig.\ref{fig:iwns class}, because these boundaries have no effect on the mode transformation with a fixed finite $k_{\|}$. 
Only the boundaries between type 1a and 1b, and type 2a and 2b are kept in the kinetic CMA diagram.
The $P+L=0,P<0$ line is the boundary for the cold plasma L wave IWNS where two inflection points on the IWNS coincide at $\theta=0$ (see Fig.\ref{fig:cold iwns surface}). 
The kinetic corrections along the $P+L=0,P<0$ line are only significant when the frequency approaches the ion cyclotron frequency. 
For most of the points on the $P+L=0, P<0$ line, the frequency is far from the ion cyclotron frequency, so the kinetic corrections are not crucial. 
In this case, $P+L=0, P<0$ is still a suitable boundary between type 1a and 1b in the kinetic CMA diagram.
The same is true for $P+R=0, P<0$, which is only inaccurate when the frequency is close to the electron cyclotron frequency.

The representative IWNSs in each region of the kinetic CMA diagram are illustrated in Fig.\ref{fig:thermal iwns surface}.
It is necessary to emphasize that the regions labelled (N) in Fig.\ref{fig:cold iwns surface} and \ref{fig:thermal iwns surface} do not represent equivalent physical regions, since the inclusion of finite-temperature kinetic effects modifies the partitioning of the parameter space.
When calculating the IWNSs, the kinetic dispersion relation is regarded as a meromorphic function of $k_{\perp}\rho_{ti}$ with fixed $\omega/\Omega_{ci}, k_{\|}\rho_{ti}$ and $\omega^{2}_{pe}/\Omega^{2}_{ce}$. 
It should be noted that there are heavily damped modes connected to type 1b (2b) IWNS extreme points in Fig.\ref{fig:thermal iwns surface}, which are discussed in detail in the Appendix B.

By comparing respectively Fig.\ref{fig:cold cma diagram} with Fig.\ref{fig:kinetic cma diagram}, and Fig.\ref{fig:cold iwns surface} with Fig.\ref{fig:thermal iwns surface}, we can clearly summarize the differences between the kinetic CMA diagram and the cold CMA diagram as follows:

\begin{figure}
  \centering
  \includegraphics[width=1.0\textwidth]{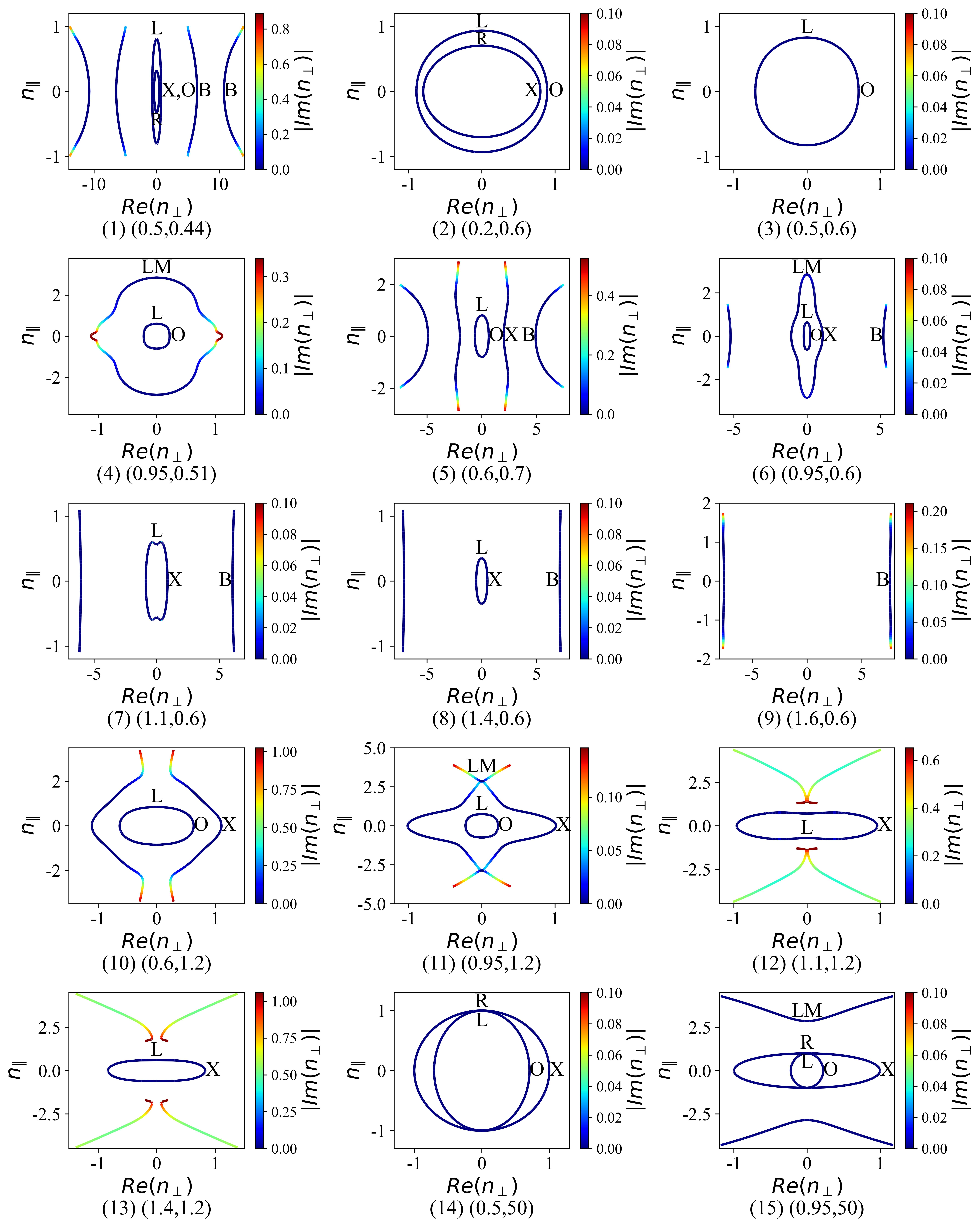} 
  \caption*{\fontsize{15}{15}(1)-(15)} 
  \label{fig:subfig11.1}
\end{figure}

\begin{figure}
  \centering
  \includegraphics[width=1.0\textwidth]{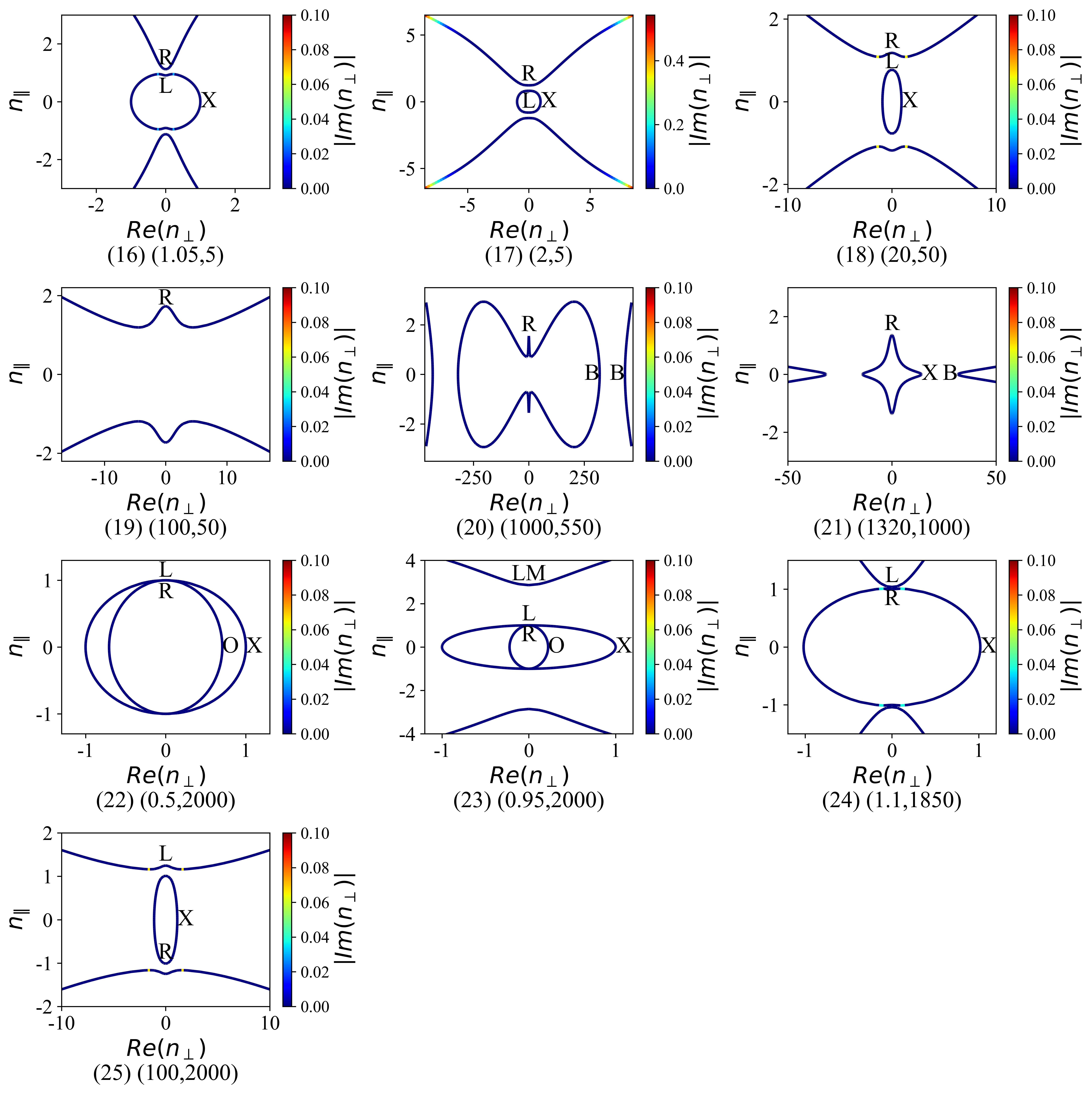} 
  \caption*{\fontsize{15}{15}(16)-(25)} 
  \label{fig:subfig11.2}
\end{figure}

\begin{figure}
  \centering
  \caption{\fontsize{12}{12}The representative IWNSs in each region of the kinetic CMA diagram, with coordinates provided below each subfigure. The color of curves indicates $Im(n_{\perp})$. The symbol B and LM represent the Bernstein wave and the Langmuir wave, respectively. In some subfigures (such as subfigure (2)), the modes are weakly damped with $|Im(n_{\perp})|$ close to 0. It is important to note that the region (N) shown in Fig.\ref{fig:cold iwns surface} and Fig.\ref{fig:thermal iwns surface} does not represent the same physical region.}
  \label{fig:thermal iwns surface}
\end{figure}

(1)Wave boundary lines (excluding the cutoff lines).
\begin{itemize}
  \item L and R waves. In the cold CMA diagram, the boundary of the L (R) wave is the ion (electron) cyclotron frequency line.
  In the kinetic CMA diagram, there are new boundaries for the weakly damped L and R waves, which are located above the ion and electron cyclotron frequency lines, respectively.
  \item Langmuir wave. In the cold CMA diagram, the Langmuir wave exists on the $\omega_{pe}^{2}/\omega^{2}=1$ line.
  In contrast, in the kinetic CMA diagram, the weakly damped Langmuir wave occupies a specific region, which is between the new Langmuir wave boundary (a vertical yellow line in Fig.\ref{fig:kinetic cma diagram}) and the $\omega_{pe}^{2}/\omega^{2}=1$ line.
  \item X waves and Bernstein waves. In the cold CMA diagram, the lower and upper hybrid resonant frequency lines serve as the boundaries of X waves.
  In the kinetic CMA diagram, the X-B mode transformation frequency lines replace the hybrid resonant frequency lines and act as the boundaries of X waves and Bernstein waves.
  Here it is worth mentioning that the X-B mode transformation frequency lines are discontinuous between different cyclotron harmonics.
\end{itemize}

Overall, these new boundaries in the kinetic CMA diagram partition the parameter space in a different manner compared to the cold CMA diagram, leading to different wave characteristics.

(2)IWNSs.
\begin{itemize}
  \item Similar regions in different diagrams. Consider Fig.\ref{fig:cold iwns surface} (3) and Fig.\ref{fig:thermal iwns surface} (5), which are located in similar regions within their respective CMA diagrams, we find Fig.\ref{fig:thermal iwns surface} (5) exhibits three branches. Among these, the X wave branch undergoes significant damping when the absolute value of the parallel refractive index $|n_{\|}|$ is large.
  On the contrary, Fig.3 (3) has two undamped branches.
  \item Same coordinate regions with different wave behaviors. Figure \ref{fig:cold iwns surface} (4) and Fig.\ref{fig:thermal iwns surface} (6) have the same coordinate in the CMA diagram. 
  However, in Fig.\ref{fig:thermal iwns surface} (6), there is an Langmuir-extraordinary (LM-X) branch.
  Along this branch, as the propagation angle varies from $\pi/2$ to 0, the X wave can continuously transform into the Langmuir wave.
  While, the X wave branch can not smoothly transform into a parallel wave in Fig.3 (4).
  \item Damping effects in specific regions. In the region between the R wave boundary and the electron cyclotron frequency line, the kinetic CMA diagram shows different wave behaviors. The R wave is heavily damped, and as a result, the outer branch of the IWNS no longer terminates at the R wave. This can be clearly seen in subfigures (10), (11), (12), and (13) of Fig.\ref{fig:thermal iwns surface}. 
  Additionally, in the region near the cyclotron harmonic frequency line, cyclotron damping phenomena are quite obvious.
  Furthermore, when the frequency is lower than the left hand cutoff frequency and the electron cyclotron frequency but higher than the R wave boundary frequency, all waves experience strong damping. 
\end{itemize}

In conclusion, these differences in IWNSs play a crucial role in understanding different wave propagation paths between the thermal plasma and cold plasma cases.
Moreover, the newly discovered IWNS characteristics have the potential to inspire the exploration of new wave propagation paths.

\subsection{The Application of The Kinetic CMA Diagram}
\begin{figure*}
  \centering
  \subfigure[$n_{\|}=2.0$]
  {
    \begin{minipage}[b]{1.0\textwidth}
      \centering
      \includegraphics[width=\textwidth]{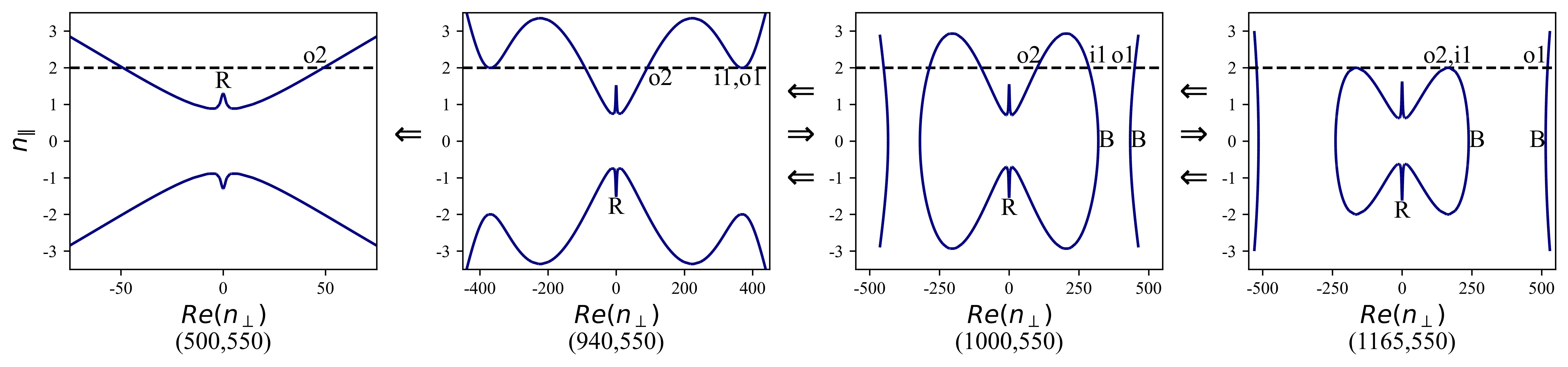}
    \end{minipage}
  }
  \subfigure[$n_{\|}=0.64$]
  {
    \begin{minipage}[b]{1.0\textwidth}
      \centering
      \includegraphics[width=\textwidth]{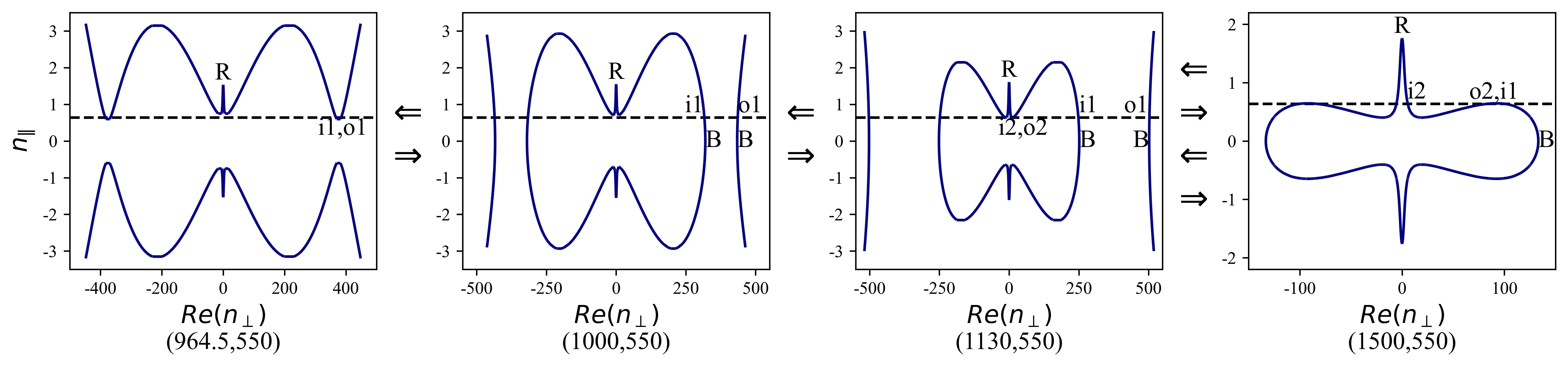}
    \end{minipage}
  }
  \subfigure[$n_{\|}=2.0$]
  {
    \begin{minipage}[b]{.45\textwidth}
      \centering
      \includegraphics[width=\textwidth]{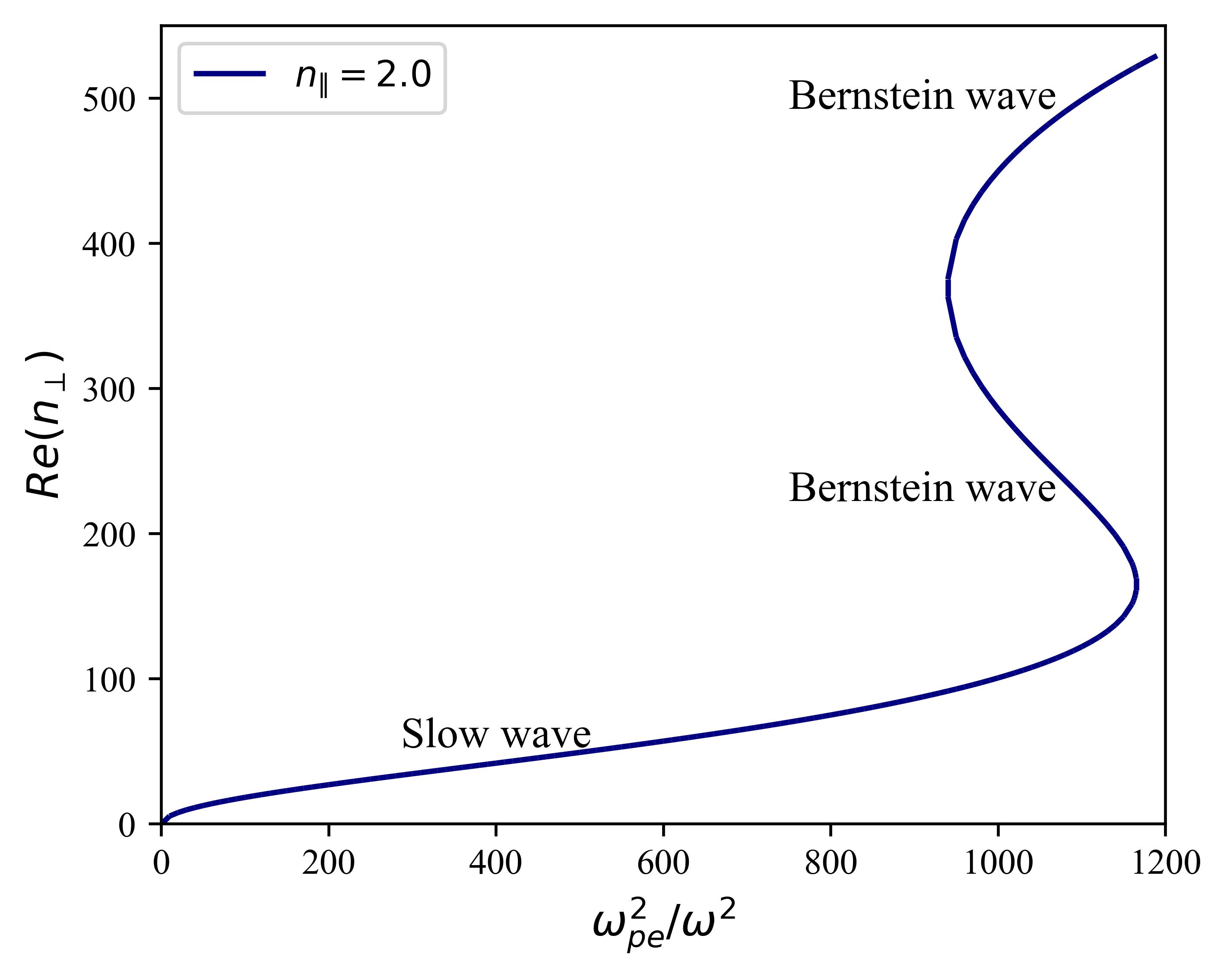}
    \end{minipage}
  }
  \subfigure[$n_{\|}=0.64$]
  {
    \begin{minipage}[b]{.45\textwidth}
      \centering
      \includegraphics[width=\textwidth]{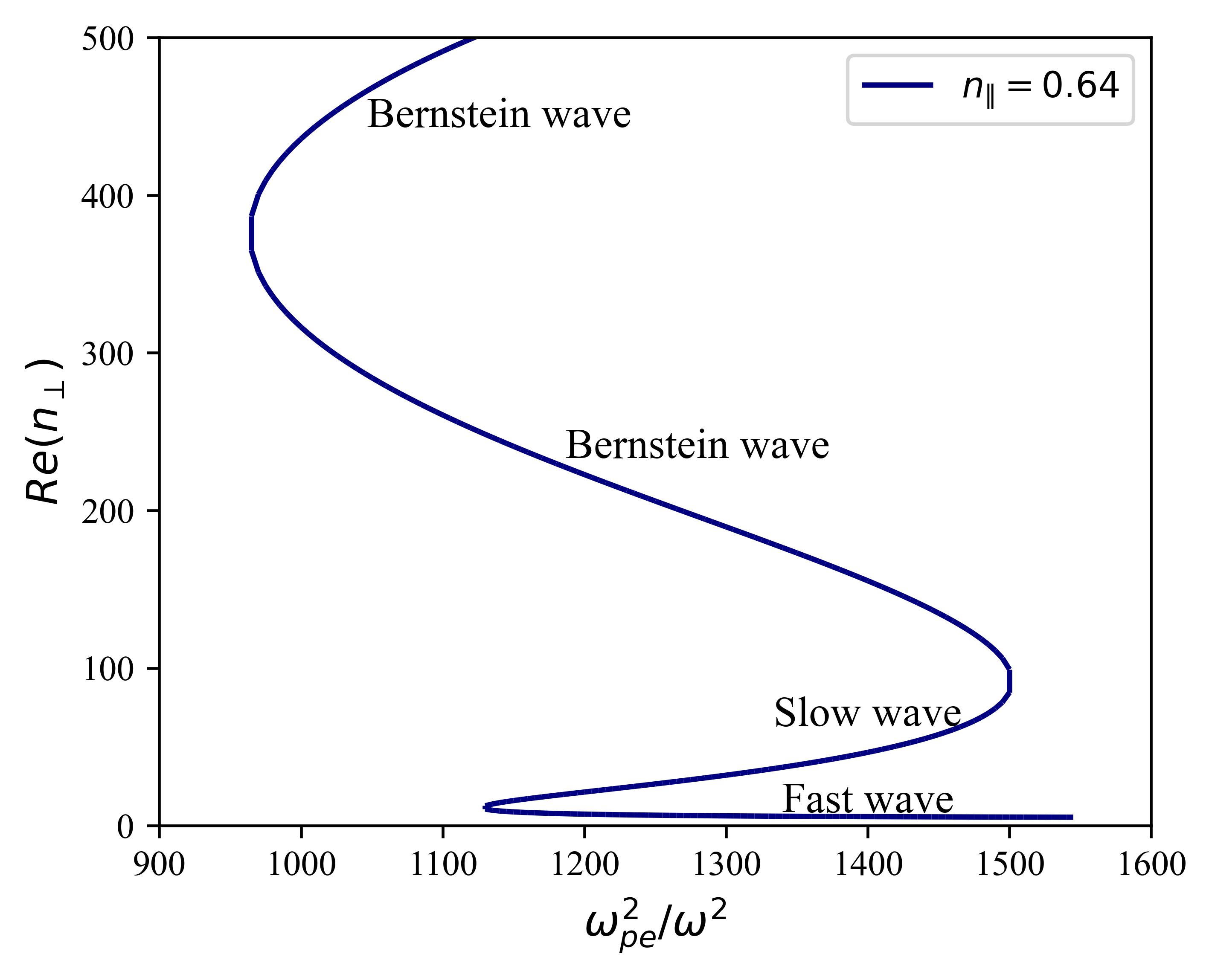}
    \end{minipage}
  }
  \caption{\label{fig:propagation paths 2} Figures (a) and (b) depict the IWNSs on the wave propagation paths in $T_{i}=T_{e}=1keV$ thermal plasma with $\Omega_{ce}/\omega=550$,$n_{\|}=2.0$ and $n_{\|}=0.64$. Figures (c) and (d) display the $n_{\perp}$ versus $\omega_{pe}^{2}/\omega^{2}$ diagrams for the same waves represented in Figures (a) and (b). It is important to compare these figures with Fig.\ref{fig:propagation paths 1} , where similar symbols are used to denote corresponding features. Specifically, "o1" and "o2" respectively represent the outward wave before the first reflection and after the second reflection.}
\end{figure*}
\begin{figure}
  \subfigure[High field side]
  {
    \begin{minipage}[b]{0.5\textwidth}
      \centering
      \includegraphics[width=\textwidth]{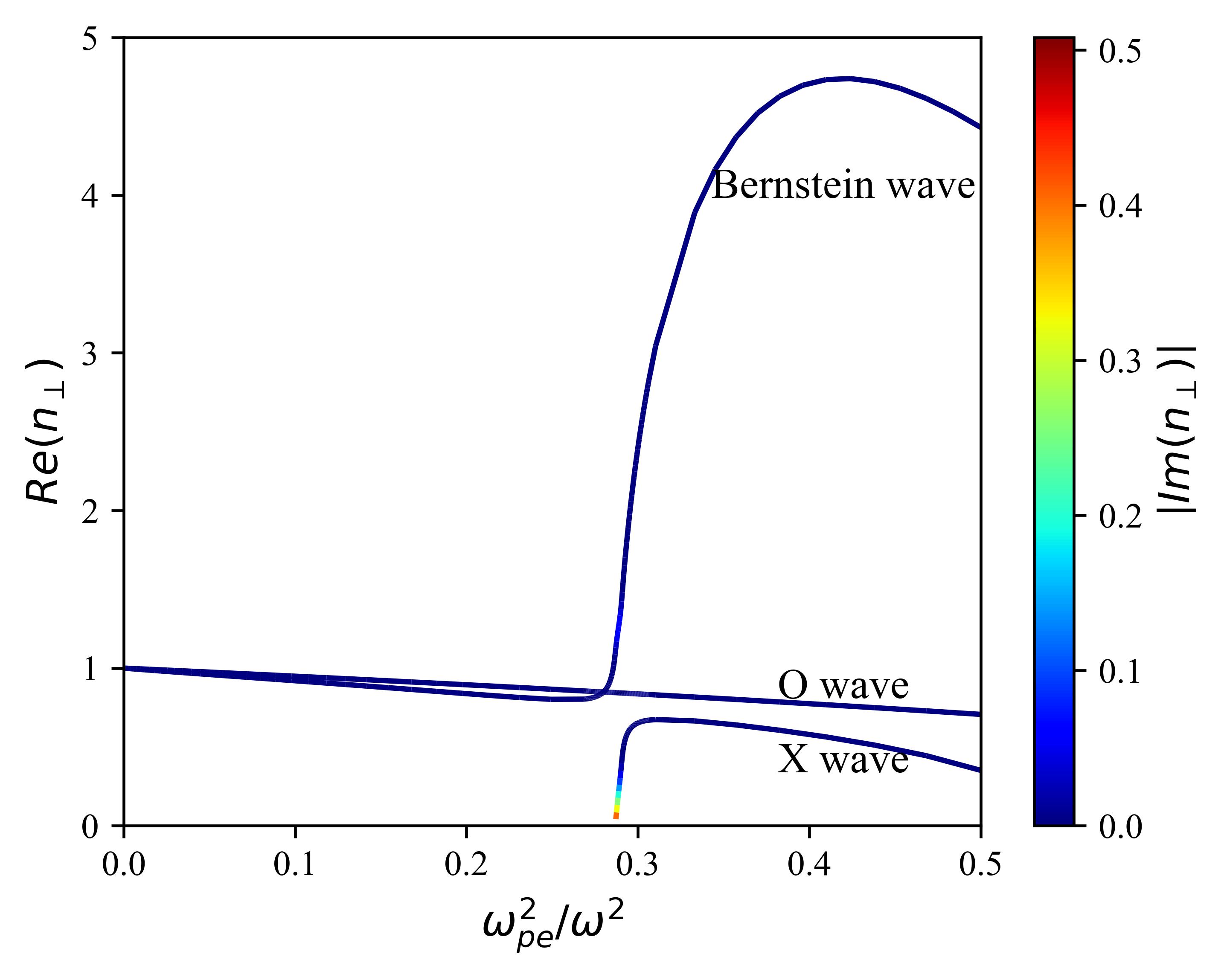}
    \end{minipage}
  }
  \subfigure[Low field side]
  {
    \begin{minipage}[b]{0.5\textwidth}
      \centering
      \includegraphics[width=\textwidth]{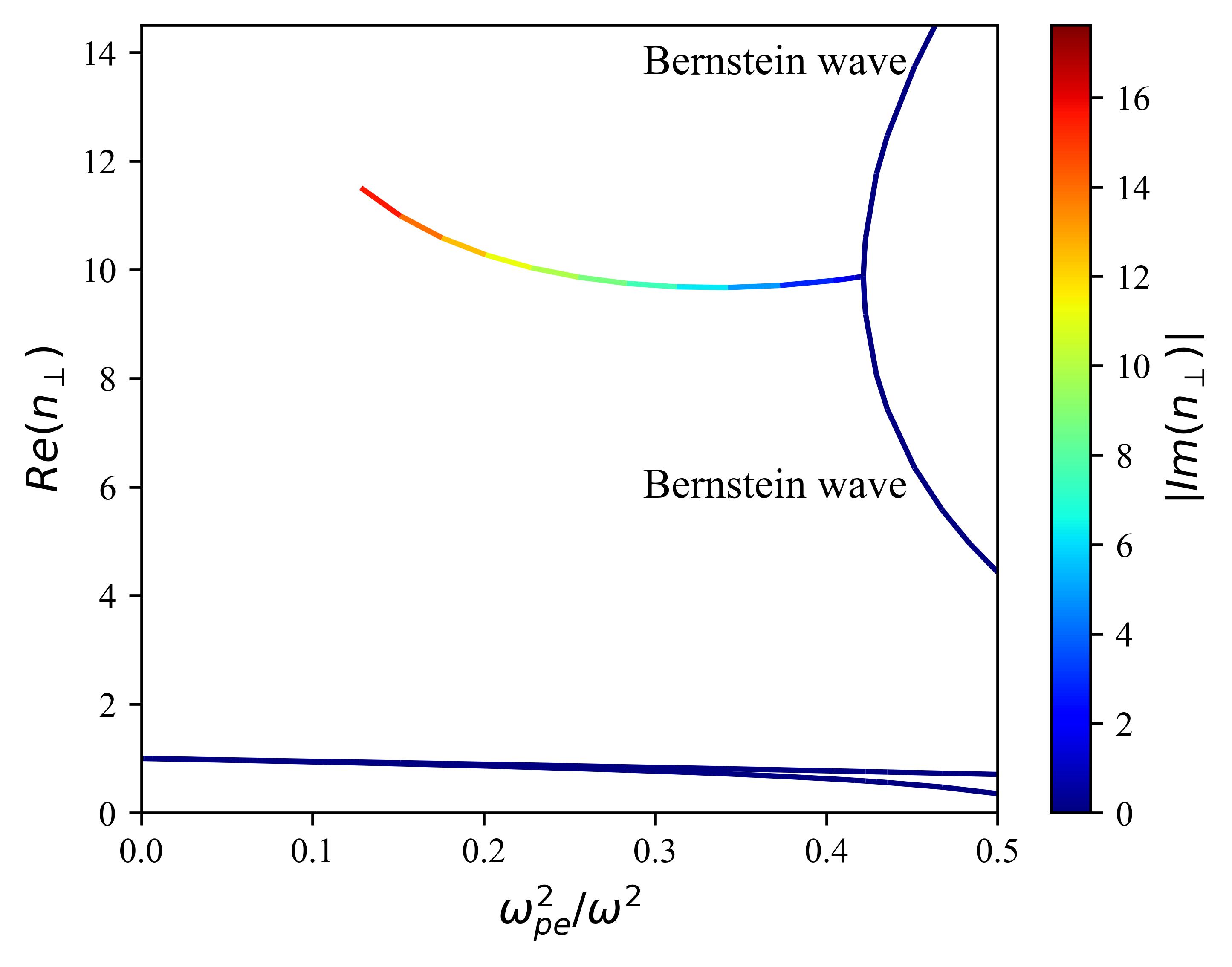}
    \end{minipage}
  }
  \caption{\label{fig:electron bernstein wave propagation}  The $n_{\perp}$ versus $\omega_{pe}^{2}/\omega^{2}$ diagrams for the electron Bernstein wave propagation paths with $n_{\|}=0.01$. In Figure (a), the high field side path is depicted, while Figure (b) displays the low field side path. These diagrams are based on the assumption that the magnetic field has the form  $B=B_{0}R_{0}/(R_{0}+r)$ and the plasma density and temperature vary as $n=n_{0}(1-|r/a|)^{2},T=T_{0}(1-|r/a|)^{2}$. Here, $B_{0},n_{0},T_{0}$ represent the physical quantity at the magnetic axis, and $r\in [-a,a]$ represents the distance from the magnetic axis. The wave propagation paths on the CMA diagram can be described by the equation $X=X_{0}(1-\left|Y_{0}/Y-1\right|R_{0}/a)^{2}$, where $(X_{0},Y_{0})=(0.5,0.46)$ denotes the starting point coordinates, and the aspect ratio $R_{0}/a=3$.}
\end{figure}

The kinetic CMA diagram provides a basic physical tool to understand the wave propagation in thermal plasmas, and it could suggest new wave propagation paths.
For example, here we present two applications about the propagation paths of Bernstein waves generated in the tokamak plasma.

1. Ion Bernstein wave

When the frequency is close to the harmonic cyclotron frequency $\omega\approx n\Omega_{ci}$, the mode transformation between the electron plasma wave and the ion Bernstein wave takes place, which has a threshold for the ion temperature near $\omega\approx \omega_{pi}$ in the transformation region \cite{Ono1993}.
The electron plasma wave exists in a very low density regime $\omega_{pi}<\omega$ and has the dispersion relation $n^{2}_{\perp}=(\omega_{pe}^{2}/\omega^{2}-1)(n^{2}_{\|}-1)$.
In this case, the ion Bernstein wave can transform into the electron plasma wave and reach the regime of $\omega_{pe}^{2}/\omega^{2}=1$.

If the frequency is not close to the harmonic cyclotron frequency, it is observed that the ion Bernstein wave can reach the plasma edge when $n_{\|}$ exceeds a specific threshold.
By comparing Fig.\ref{fig:propagation paths 1} (a) and Fig.\ref{fig:propagation paths 2} (a), we can see that when the frequency is near the X-B mode transformation frequency in the kinetic CMA diagram, the R wave IWNS  experiences topological changes.
This topological change implies that when the constant $n_{\|}$ line is tangent to the R wave IWNS at the maximum $|n_{\|}|$ point, the slow wave can convert into the Bernstein wave.
For example, Fig.\ref{fig:propagation paths 2} (a) and (c) illustrate the propagation path for a slow wave with $n_{\|}=2.0,{\Omega_{ce}}/{\omega}=550,T_{i}=1keV$.
Different from the cold plasma case, in this scenario, the slow wave is not confined by the lower hybrid resonant frequency line.
Instead, it can transform into the Bernstein wave and propagate through the higher density plasma \cite{stix1965}.
This wave propagation in the opposite direction suggests that in tokamak plasmas, when $n_{\|}$ is above a certain threshold, the ion Bernstein wave can propagate to the plasma edge (${\omega_{pe}^{2}}/{\omega^{2}}=1$) and might be measured outside the plasma.
When $n_{\|}$ is below the threshold, the slow wave, which is converted from the Bernstein wave, undergoes reflection and transformation into the fast wave, as illustrated in Fig.\ref{fig:propagation paths 2} (b)and (d).
The difference between these two propagation paths lies in the slow-fast wave mode transformation.
The slow-fast wave mode transformation occurs in the regime where $k_{\perp}\rho_{ti}\ll 1$ and is minimally affected by the finite Larmor radius effect.
Therefore, using the cold plasma results of Eq.(\ref{eq:n_min}), we can approximate $n_{\|,c}$ as $n_{\|,c}\simeq max\left\{\sqrt{{S(P-R)(P-L)}/{(P-S)^2}}+\sqrt{{-PD^{2}}/{(P-S)^2}}\right\}$ .
In the illustrated case of ${\Omega_{ce}}/{\omega}=550$, the value of $n_{\|,c}$ is 1.003.

2. Electron Bernstein wave

It is found that measuring the electron Bernstein wave on the high field side may be more feasible than on the low field side. 
The X-B mode transformation lines depicted in Fig.\ref{fig:kinetic cma diagram} exhibit discontinuities among different cyclotron harmonics. 
This observation suggests the existence of a wave propagation path that intersects the cyclotron harmonic frequency line and exits the Bernstein wave region.
As a practical example, we consider a scenario in which an electron Bernstein wave exists within the tokamak plasma, specifically located in region (1) of Fig.\ref{fig:kinetic cma diagram}. 
In this case, the wave's frequency is higher than the right-hand cutoff frequency and the double cyclotron harmonic frequency, but lower than the X-B mode transformation frequency. 
The red and blue arrows originating from region (1) in Fig.\ref{fig:kinetic cma diagram} (b) represent the propagation paths of the Bernstein wave toward the high field side and the low field side, respectively. 
Fig.\ref{fig:electron bernstein wave propagation} illustrates the detailed dependence of $n_{\perp}$ on $\omega_{pe}^{2}/\omega^{2}$, with Fig.\ref{fig:electron bernstein wave propagation} (a) showing the path of the Bernstein wave on the high field side and Fig.\ref{fig:electron bernstein wave propagation} (b) depicting the path on the low field side.
When the Bernstein wave propagates toward the low field side, it encounters the X-B mode transformation line, resulting in its reflection back into a higher density region.
In contrast, the Bernstein wave travels a greater distance along the high field side path—from the starting point to the double cyclotron harmonic line—than it does along the low field side path, as illustrated by the red arrow in Fig.\ref{fig:kinetic cma diagram} (b).
Further calculations suggest that upon crossing the double cyclotron harmonic line, the Bernstein wave may transform into an X wave and reach the plasma edge, as shown in Fig.\ref{fig:electron bernstein wave propagation} (a).
It is crucial to emphasize that this phenomenon is specific to the Bernstein wave, whose frequency range lies between the X-B mode transformation frequency and the right-hand cutoff frequency.

\subsection{Discussions}
The kinetic CMA diagram in section 3.3 assumes a fixed temperature $T=1keV$.
To investigate the influence of temperature on the kinetic CMA diagram, Fig.\ref{fig:kinetic cma diagram 3d} introduces the temperature as an additional z axis.
To avoid visual overlap between surfaces and improve clarity, the diagram is restricted to the regimes of $0<\omega_{pe}^{2}/\omega^{2}<2$, $0<\Omega_{ce}/\omega<2$ and only one X-B mode transformation surface between $\Omega_{ce}/\omega=0.5$ and $\Omega_{ce}/\omega=1$ is demonstrated.
As shown in Fig.\ref{fig:kinetic cma diagram 3d}, the X-B mode transformation line converges to the upper hybrid resonant line ,and the R wave boundary approaches the electron cyclotron frequency line as the temperature decreases from 1keV to 0keV.
It should be noted that the Langmuir wave boundary, which is defined by the frequency criterion $\omega_{i}=-0.01\omega_{r}$, exhibits no temperature dependence on the $(\omega^{2}_{pe}/\omega_{r}^{2},\Omega_{ce}/\omega_{r})$ plane.
This can be seen from the dispersion relation of the Langmuir wave
\begin{equation}
    1+2\sum_{\alpha=i,e}\frac{\omega_{p\alpha}^2}{k^2_{\|}v_{t\alpha}^2}\left[1+\frac{\omega}{k_{\|}v_{t\alpha}}Z\left(\frac{\omega}{k_{\|}v_{t\alpha}}\right)\right]=0,
    \label{eq:Langmuir}
\end{equation}
where $\omega_{p\alpha}$ is the plasma frequency of species $\alpha$, $v_{t\alpha}$ is the thermal velocity of species $\alpha$ $\left(v_{t\alpha}=\sqrt{2T_{\alpha}/{m_{\alpha}}}\right)$, and $Z(x)$ is the plasma dispersion function.
When the frequency $\omega$ satisfies the condition $\omega_{i}=-0.01\omega_{r}$, Eq.(\ref{eq:Langmuir}) contains three real variables $(\omega_{r},k_{\|}v_{t\alpha},\omega_{p\alpha})$ but is constrained by two independent equations.
When $\omega_{p\alpha}$ is given, both $\omega_{r}$ and $k_{\|}v_{t\alpha}$ can be determined.
Consequently, the Langmuir wave boundary, which is plotted on the $(\omega^{2}_{pe}/\omega_{r}^{2},\Omega_{ce}/\omega_{r})$ plane in the kinetic CMA diagram, exhibits no temperature dependence.
The temperature only affects the IWNSs of the Langmuir wave.

Furthermore, temperature variations introduce new regions in the kinetic CMA diagram. 
For example, the X-B mode transformation surface has a intersection line with the Langmuir wave boundary surface. 
Therefore, a region enclosed by the X-B mode transformation surface, the Langmuir wave boundary surface and the electron harmonic cyclotron frequency surface will appear when the temperature is above a certain threshold, denoted as region (4) in Fig.\ref{fig:kinetic cma diagram}.
In this region, the Langmuir wave can not smoothly transform into a X wave as the propagation angle changes from 0 to $\pi/2$.

\begin{figure}
    \centering
    \includegraphics[width=1.0\textwidth]{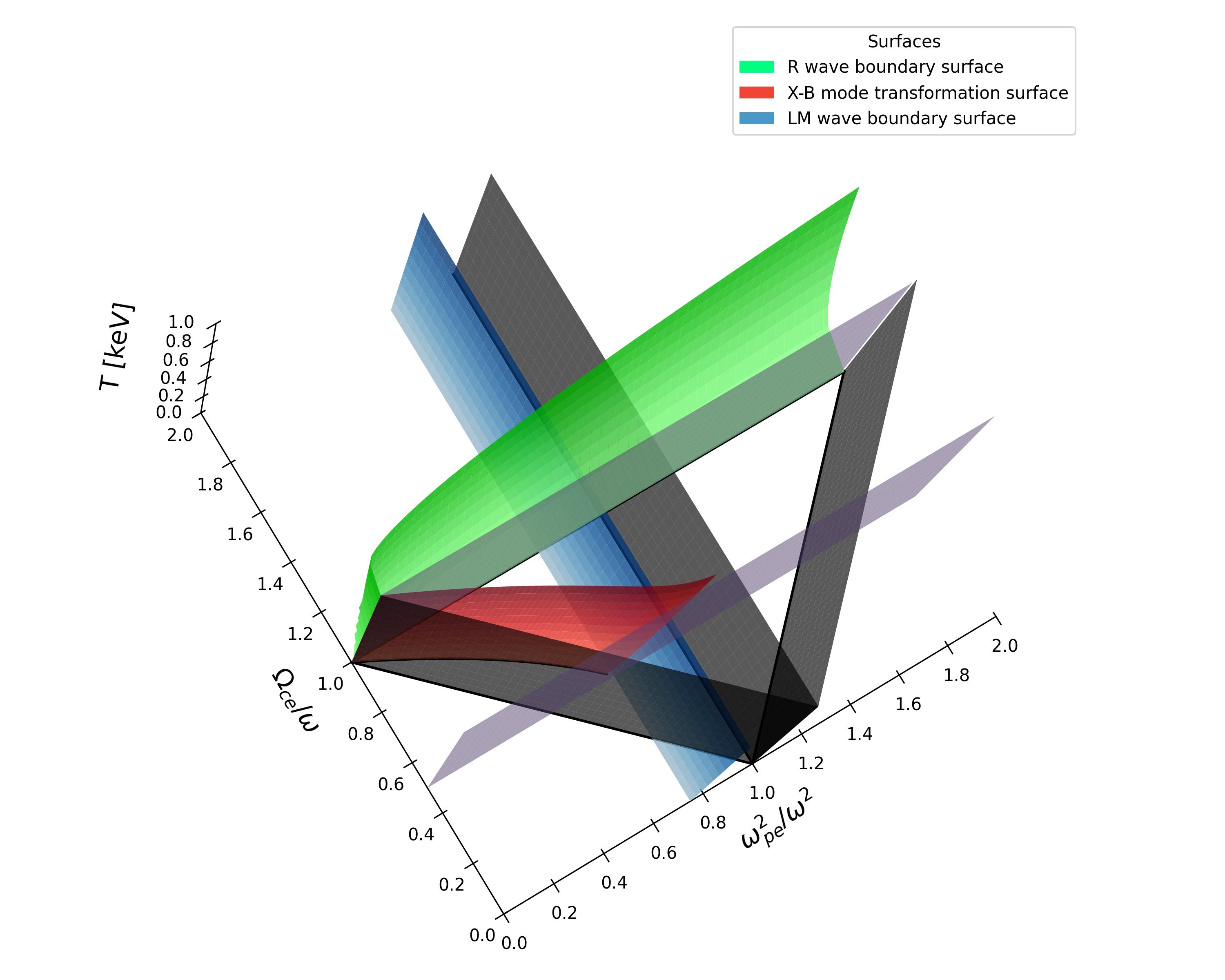} 
    \caption*{{(a) Top View $(T=1keV\rightarrow 0keV)$}}
    \label{fig:subfig1}
\end{figure}

\begin{figure}
    \centering
    \includegraphics[width=1.0\textwidth]{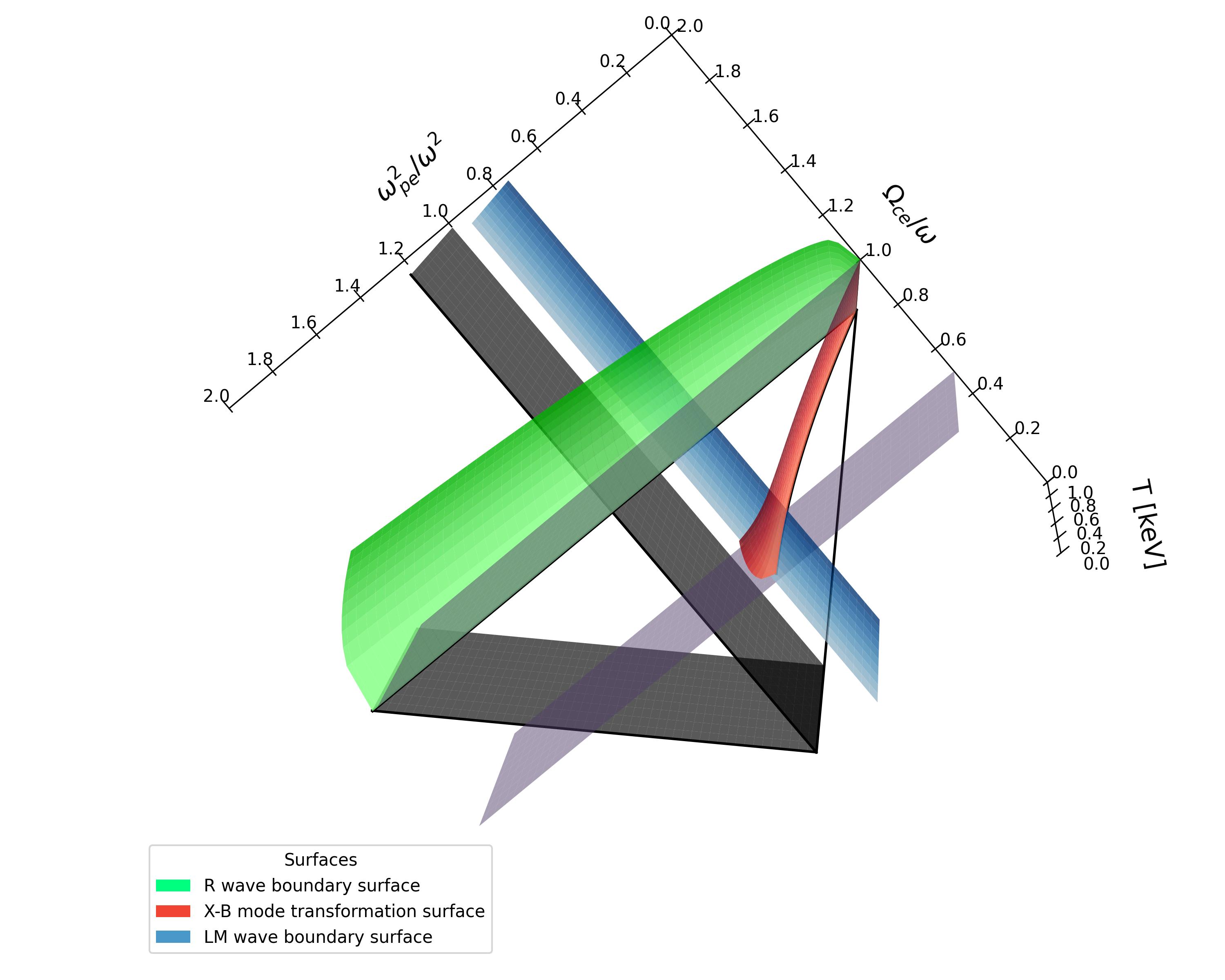} 
    \caption*{{(b) Bottom View $(T=0keV\rightarrow 1keV)$} }
    \label{fig:subfig2}
\end{figure}

\begin{figure}
    \centering
    \caption{ The kinetic CMA diagram with the temperature as the z axis. 
    Subfigure (a) is the top view of the kinetic CMA diagram where the temperature on the z axis decreases from $1keV$ to $0keV$.
    Similarly, subfigure (b) is the bottom view with the temperature axis increasing from $0keV$ to $1keV$.
    For clarity, only the regimes specified by $0<\omega_{pe}^{2}/\omega^{2}<2$ and $0<\Omega_{ce}/\omega<2$ are displayed. 
    The green, red, and blue surfaces respectively denote the R wave boundary surface, the X-B mode transformation surface, and the Langmuir wave boundary surface, with the color intensity on these surfaces indicating the temperature magnitude. 
    The black surfaces signify cutoff surfaces, including the plasma frequency surface, the right- and left-hand cutoff surfaces. 
    Meanwhile, the light purple surfaces represent the electron cyclotron harmonic surfaces. 
    Significantly, the intersection lines of these surfaces with the $T=0keV$ plane are outlined in black, thereby depicting the cold CMA diagram.}
    \label{fig:kinetic cma diagram 3d}
\end{figure}

\begin{figure*}
  \centering
  \subfigure[The CMA diagram for $1836<\Omega_{ce}/\omega<4000$]
  {
    \begin{minipage}[b]{.7\textwidth}
      \centering
      \includegraphics[width=\textwidth]{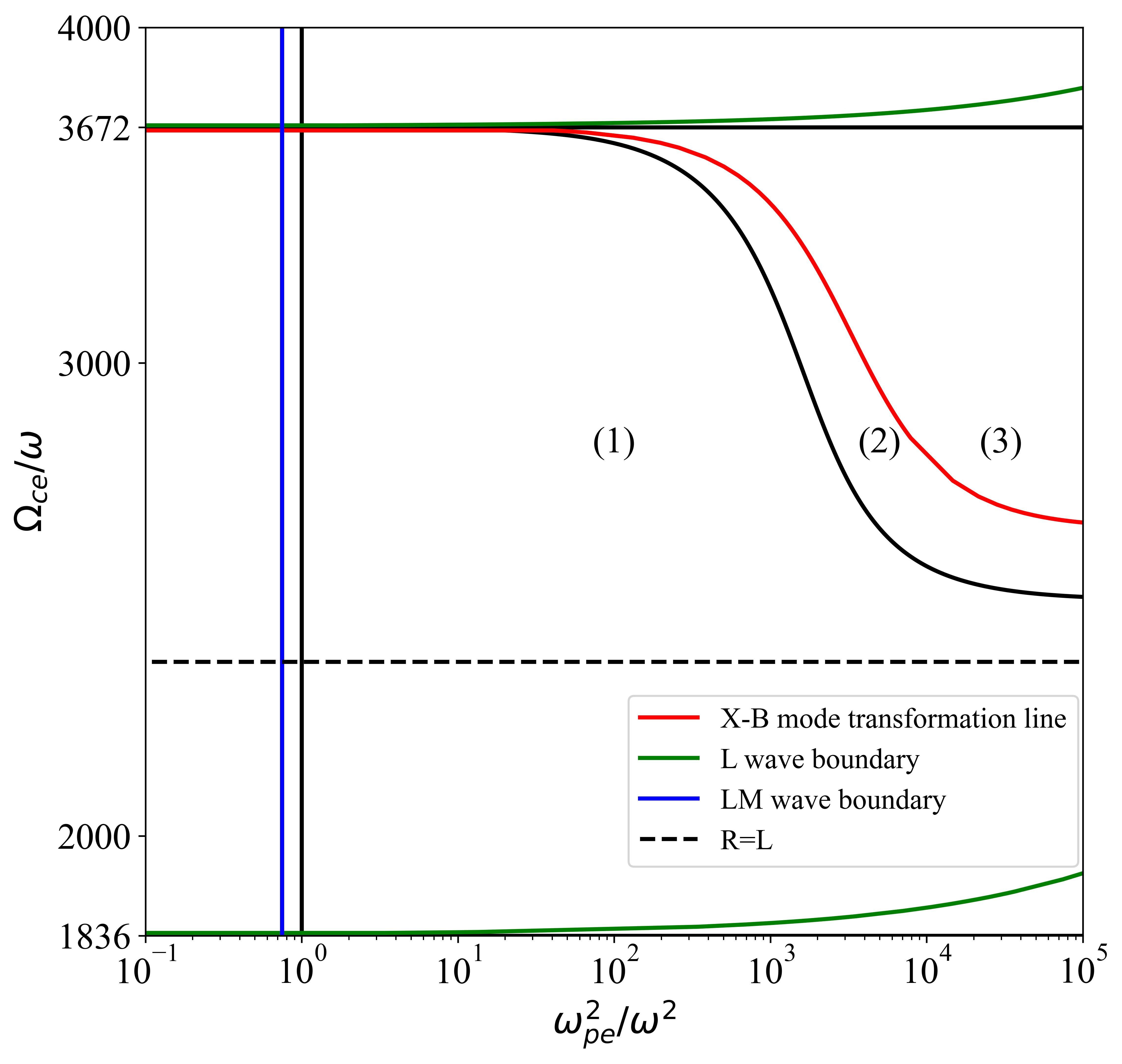}
    \end{minipage}
  }
  \subfigure[Representative IWNSs]
  {
    \begin{minipage}[b]{1.0\textwidth}
      \centering
      \includegraphics[width=\textwidth]{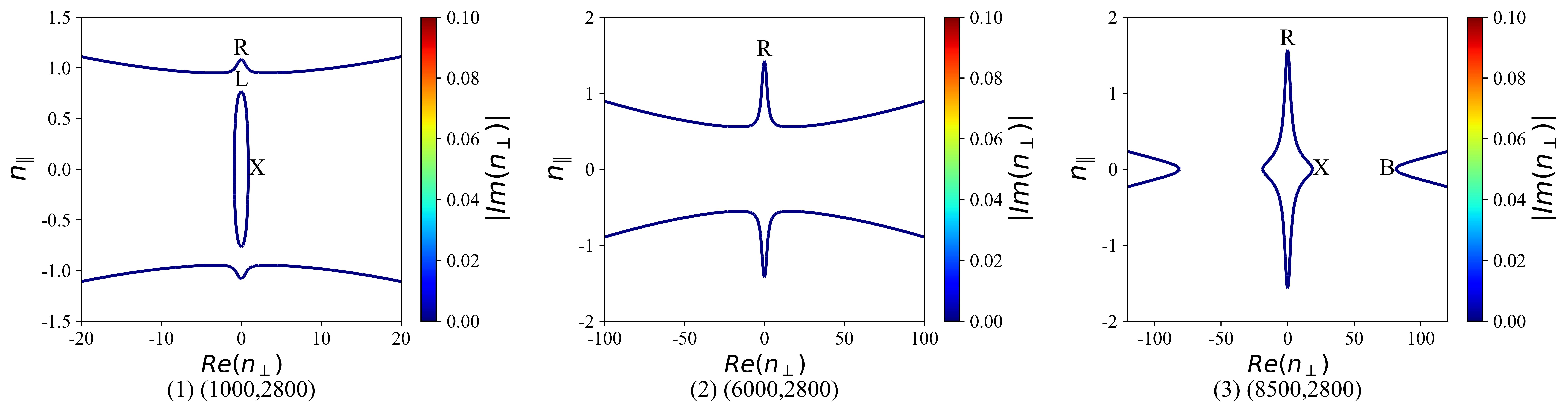}
    \end{minipage}
  }
  \caption{\label{fig:two ion species cma diagram}  The kinetic CMA diagram for two ion species plasma. Subfigure (a) shows the kinetic CMA diagram in the regime of $1836<\Omega_{ce}/\omega<4000$. Meanwhile, subfigure (b) displays the representative IWNSs in this kinetic CMA diagram.}
\end{figure*}

The discussion up to this point has focused on the kinetic CMA diagram for a single ion species plasma.
To investigate the influence of the second ion species on the kinetic CMA diagram, we consider a plasma consisting of hydrogen ions and deuterium ions with equal density and temperature.
In the scenario of cold plasma, when the second ion species is included, a new left hand cutoff frequency and an ion-ion hybrid resonant frequency emerge in the regime of $0<\omega/\Omega_{ci}<1$ \cite{stott,walker3}. 
Analogous to the single ion species plasma case, the FLR effect modifies the ion-ion hybrid resonant frequency to the X-B mode transformation frequency.
As an example, Fig.\ref{fig:two ion species cma diagram} (a) illustrates the kinetic CMA diagram for the two ion species plasma in the regime of $1836<\Omega_{ce}/\omega<4000$.
Meanwhile, Fig.\ref{fig:two ion species cma diagram} (b) demonstrates the representative IWNSs in the region bounded by the new left hand cutoff frequency line and X-B mode transformation frequency line.
It can be seen that the IWNSs in the region (1), (2) and (3) of Fig.\ref{fig:two ion species cma diagram} are similar to those in the region (18), (19) and (21) of the kinetic CMA diagram for the single ion species plasma (Fig.\ref{fig:kinetic cma diagram}), respectively.
Futhermore, the inclusion of the second ion species also introduces new X-B mode transformation frequency lines within the cyclotron harmonic frequency range of the second ion.
More detailed analysis of the kinetic CMA diagram for two ion species plasma will be presented in future work.

\section{Conclusions and Discussions}
In this study, we introduce a kinetic CMA diagram for thermal plasmas incorporating kinetic effects, such as the FLR effect, Landau damping and cyclotron damping.
By using a new eigenvalue solver, we calculate the new wave boundaries and the representative IWNSs in each region, which combine to build the kinetic CMA diagram framework.
The kinetic effects on wave propagation are systematically analysed.

There are significant differences between the cold and kinetic CMA diagrams.
In terms of wave boundary lines, the kinetic CMA diagram presents new boundaries. 
For L and R waves, their new boundaries are above the ion and electron cyclotron frequency lines respectively, unlike in the cold CMA diagram.
The Langmuir wave occupies a specific region between a new boundary and the $\omega_{pe}^{2}/\omega^{2}=1$ line in the kinetic CMA diagram, rather than just existing on the $\omega_{pe}^{2}/\omega^{2}=1$ line.
The X-B mode transformation frequency lines replace the hybrid resonant frequency lines as boundaries for X and Bernstein waves, with discontinuities between cyclotron harmonics. 
These new boundaries partition the parameter space differently, resulting in different wave behaviors.
Regarding IWNSs, even when considering the same coordinates in the two diagrams, the number and topological structure of wave branches can be different, particularly in the regions defined by the new boundaries.
In regions surrounded by the new boundary of R (L) wave and electron (ion) cyclotron frequency line, damping effects are quite prominent.
These differences in IWNSs are important for understanding different wave propagation paths in both thermal and cold plasmas.

The kinetic CMA diagram can provide a comprehensive framework for investigating wave propagation in thermal plasmas.
Considering the tokamak plasma as an application, it is found that measuring the electron Bernstein wave on the high field side may be more feasible than on the low field side.
The X-B mode transformation lines in the kinetic CMA diagram have discontinuities between different cyclotron harmonics, indicating a possible wave propagation path intersecting the cyclotron harmonic frequency line and leaving the Bernstein wave region.
When the electron Bernstein wave is in region (1) of Fig.\ref{fig:kinetic cma diagram}, with its frequency higher than the right hand cutoff and double cyclotron harmonic frequencies but lower than the X-B mode transformation frequency, its propagation paths on the high and low field sides differ. 
On the low field side, it reflects upon reaching the X-B mode transformation line, while on the high field side, it travels a longer distance to the double cyclotron harmonic line and may transform into an X wave to reach the plasma edge. This unique behavior is specific to the electron Bernstein wave within the particular frequency range, offering new perspectives on wave propagation research in tokamak plasmas.

Finally, we remark that the present work is limited to single ion species uniform, non-relativistic, Maxwellian plasma.
It is desirable to extend the current CMA diagram to multiple ion species and relativistic plasma in the future.

\ack
This work was supported by the National MCF Energy R \& D Program of China under Grant No.2024YFE03230300, National Natural Science Foundation of China under Grant No.12375213, 12125502 and 12335014, Natural Science Foundation of Sichuan Province under Grant No.2025ZNSFSC0061, China National Nuclear Corporation ‘Young Talents’ Project No.2024-QNYC-02 and the Innovation Program of Southwestern Institute of Physics (202301XWCX001).

\appendix

\section{Derivation of the IWNS boundaries}
Here we give a brief derivation of the IWNS boundaries in the CMA diagram.
The boundary of type 1c in Fig.\ref{fig:iwns class} is omitted because it is nonessential for analysing wave propagation with fixed finite $k_{\|}$.
Starting from the cold plasma dispersion relation (Eq. \ref{eq:cold_dispersion_relation}), we substitute $tan^{2}\theta={n^{2}_{\perp}}/{n^{2}_{\|}}$ into Eq. (\ref{eq:cold_dispersion_relation}) and solve $n_{\|}^{2}$ for $P\neq S$:
\begin{eqnarray}
  n_{\|}^2&=&-\frac{S}{P-S}n^{2}-\frac{P(P-R)(P-L)\left(R L-S^2\right)}{(P-S)^3n^{2}+(P-S)^{2}(RL-PS)}\nonumber\\
  &&+\frac{P(PS+RL-2S^{2})}{(P-S)^{2}}.
  \label{eq:hook function 1}
\end{eqnarray}
For a fixed $n_{\|}^{2}$, this equation can admit 0, 1, or 2 real solutions for $n^{2}$, depending on the parameter regime.

An IWNS is categorized as type 1b or type 2b (Fig.\ref{fig:iwns class}) if Eq. (\ref{eq:hook function 1}) has an extermum point (where $\partial n^{2}_{\|}/\partial n^{2} =0$) within the physical regime $n^{2}>n^{2}_{\|}>0$ \cite{stott}.
The transition between type 1a and 1b (type 2a and 2b) occurs when the extremum coincides with the line $n^{2}=n^{2}_{\|}$.
Solving this condition yields: 

(1) $(R-L)^2(P-L)(P+L)=0,L>0$ for the L wave.

(2) $(R-L)^2(P-R)(P+R)=0,R>0$ for the R wave.

However, not all factors in these equations represent physical boundaries.

(1) The term $(R-L)^{2}\geq 0$ does not contribute to boundary transitions, as its sign reversals do not alter the extremum's position;

(2) When $P-R$ or $P-L$ crosses 0, the function described in Eq.(\ref{eq:hook function 1}) transforms from a non-monotonic to a monotonic function of $n^{2}$. 
Further numerical analyses confirm that the IWNS type remains unchanged when crossing the lines $P-R=0$ or $P-L=0$ in the CMA diagram.

Therefore, the boundaries between type 1a and 1b (type 2a and 2b) are determined by $P+L=0,P<0$ for the L wave and $P+R=0,P<0$ for the R wave.

Similarly, we can also write $n_{\perp}^{2}$ for $P \neq S$ as
\begin{eqnarray}
  n_{\perp}^2&=&\frac{P}{P-S}n^{2}+\frac{P(P-R)(P-L)\left(R L-S^2\right)}{(P-S)^3n^{2}+(P-S)^{2}(RL-PS)}\nonumber\\
  &&-\frac{P(PS+RL-2S^{2})}{(P-S)^{2}}.
  \label{eq:hook function 2}
\end{eqnarray}
To classify IWNS types, we then examine extremum points (where $\partial n^{2}_{\perp}/\partial n^{2}=0$) intersecting the line $n^{2}_{\perp}=n^{2}$ \cite{stott}.

(1)$(P-L)(P-R)=0,P>0$ for the O wave.

(2)$(R-L)^{2}\left[(P+S)RL-2PS^{2}\right]=0,{RL}/{S}>0$ for the X wave.

Again, $P-L=0$, $P-R=0$ and $R-L=0$ are not boundaries between different IWNS types.
The IWNS boundary between type 1a and 1b* (type 3a and 3b) for the X wave branch is given by $(P+S)RL-2PS^{2}=0$ and ${RL}/{S}>0$.

When $P=S$, the dispersion relation reduces to $n_{\perp}^{2}=-{4P(n^{2}-2Pn+RL)}/{(R-L)^{2}}$.
Noting that this quadratic form has an extreme point at $n^{2}=P$, which crosses the line $n^{2}=P$ as $P-S$ changes sign. 
Consequently, the boundaries between type 1a and 1b* for the O wave branch can be expressed as $P-S=0$ and $P>0$.

A summary of IWNS boundaries has been given in table.\ref{table1}.

\section{The Detailed Type 1b (2b) IWNS in The CMA Diagram}

\begin{figure}
  \subfigure[]
  {
    \begin{minipage}[b]{0.5\textwidth}
      \centering
      \includegraphics[height=0.2\textheight]{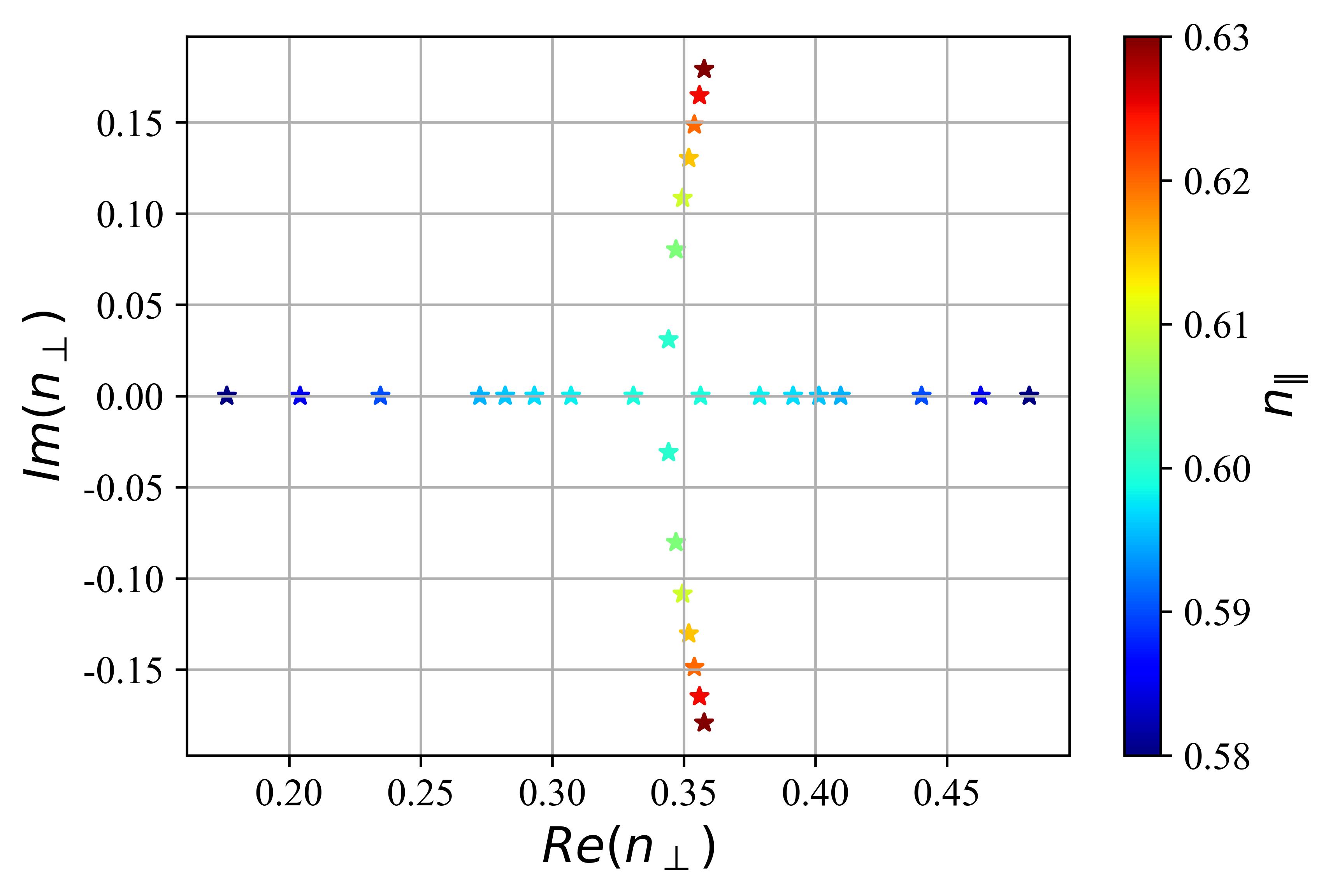}
    \end{minipage}
  }
  \subfigure[]
  {
    \begin{minipage}[b]{0.5\textwidth}
      \centering
      \includegraphics[height=0.2\textheight]{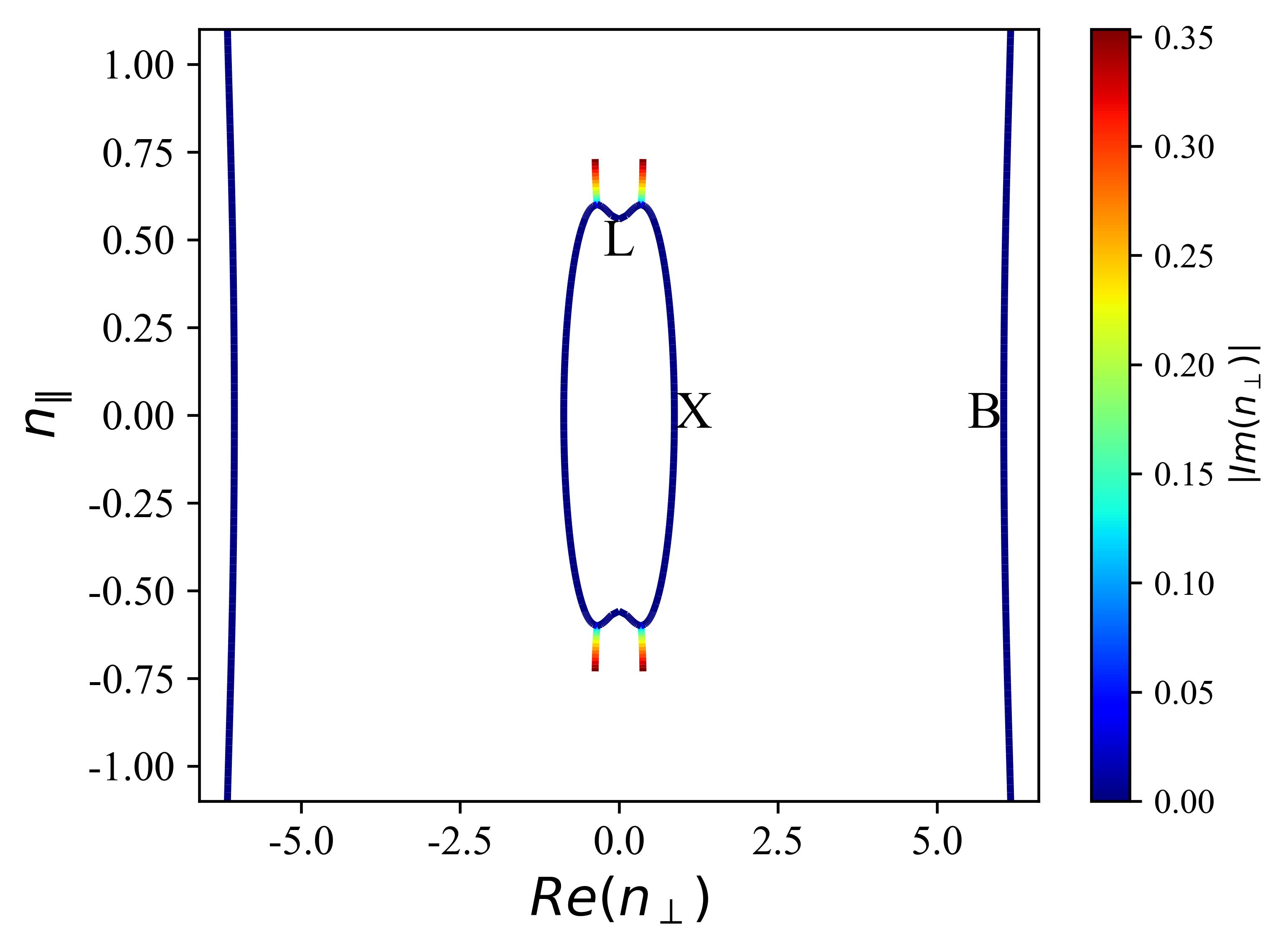}
    \end{minipage}
  }
  \caption{\label{fig:1.1_0.6_IWNS}  Figure (a) represents the relationship between  $n_{\perp}$ and $n_{\|}$ for kinetic normal modes with ${\omega_{pe}^{2}}/{\omega^{2}}=1.1,{\Omega_{ce}}/{\omega}=0.6$. The two kinetic normal modes on the real axis move in opposite directions, coincide and eventually split into one upward mode and one downward mode as $n_{\|}$ increases. Figure (b) displays the detailed type 1b IWNS of (1.1,0.6) in the kinetic CMA diagram.}
\end{figure}

For the type 1b IWNS (such as Fig.\ref{fig:thermal iwns surface}  (7)), the two modes coincide when the constant $n_{\|}=n_{\|}^{*}$ line is tangent to the IWNS at the $\theta\neq 0$ extreme point.
When $|n_{\|}|$ increases from the critical value $|n_{\|}^{*}|$, the $n_{\perp}$ solutions of the kinetic dispersion relation  split into two solutions, one moves upward while the other moves downward, as shown in Fig.\ref{fig:1.1_0.6_IWNS} (a).
This phenomenon is also observed in the cold plasma dispersion relation.
Consequently, the detailed type 1b IWNS exhibits a peak originating from the mode conversion point, as illustrated in  Fig.\ref{fig:1.1_0.6_IWNS} (b).
The type 2b IWNS displays a similar phenomenon.
To avoid confusion with other collisionless damped modes, the heavily damped modes in type 1b (2b) IWNS are removed in the text, as can be seen in Fig.\ref{fig:cold iwns surface} and Fig.\ref{fig:thermal iwns surface}.

\section*{References}

\end{document}